\documentstyle[12pt,psfig]{article}
\def\reference{\parskip 0pt\par\noindent\hangindent 0.5 truecm}
\def\kms{km ${\rm s}^{-1}$}

\def\farcs{\hbox{$.\!\!^{\prime\prime}$}}
\def\fs{\hbox{$.\!\!^{\rm s}$}}
\begin{document}
%
%
\title{Evolution of the Radio Remnant of SN 1987A: 1990 -- 2001}
%


\author{R. N. Manchester $^{1}$ \and
 B. M. Gaensler $^{2}$\thanks{Hubble Fellow. Present address: Harvard-Smithsonian Centre for Astrophysics, Cambridge MA 02138, USA} \and
 V. C. Wheaton $^{1,3}$ \and
 L. Staveley-Smith $^{1}$ \and
 A. K. Tzioumis $^{1}$ \and
 N. S. Bizunok $^{2,4}$ \and
 M. J. Kesteven $^{1}$ \and
 J. E. Reynolds $^{1}$ 
} 

\date{}
\maketitle

{\center
$^1$ Australia Telescope National Facility, CSIRO, PO Box 76, Epping NSW
1710 \\Dick.Manchester@csiro.au\\[3mm]
$^2$ Center for Space Research, Massachusetts
Institute of Technology, Cambridge MA 02139, USA \\[3mm]
$^3$ School of Physics, University of Sydney, NSW 2006\\[3mm]
$^4$ Boston University, Boston MA 02215, USA\\[3mm]
}

%
\begin{abstract}
The development of the radio remnant of SN 1987A has been followed using the
Australia Telescope Compact Array since its first detection in 1990
August. The remnant has been observed at four frequencies, 1.4, 2.4, 4.8 and
8.6~GHz, at intervals of 4 -- 6 weeks since the first detection. These data
are combined with the 843 MHz data set of Ball et al. (2001) obtained at
Molonglo Observatory to study the spectral and temporal variations of the
emission. These observations show that the remnant continues to increase in
brightness, with a larger rate of increase at recent times. They also show
that the radio spectrum is becoming flatter, with the spectral index
changing from $-0.97$ to $-0.88$ over the 11 years. In addition, at roughly
yearly intervals since 1992, the remnant has been imaged at 9~GHz using
super-resolution techniques to obtain an effective synthesised beamwidth of
about 0\farcs5.  The imaging observations confirm the shell-morphology of
the radio remnant and show that it continues to expand at $\sim3000$~\kms.
The bright regions of radio emission seen on the limb of the shell do not
appear to be related to the optical hotspots which have subsequently
appeared in surrounding circumstellar material.

\end{abstract}


\bigskip

%
%

\section{Introduction}
As the closest observed supernova in nearly 400 years, SN 1987A in the Large
Magellanic Cloud offers unique opportunities for detailed study of the
evolution of a supernova and the birth of a supernova remnant.
In the 14 years since the explosion was detected, it has been
extensively studied in all wavelength bands from radio to gamma-ray using
both ground- and space-based observatories. These observations have revealed
a complex evolution of both the supernova itself and the supernova remnant
which is developing as the ejecta and their associated shocks interact with
the circumstellar material. 

Perhaps the most dramatic of these observations are the optical observations
which show the beautiful triple-ringed structure surrounding and illuminated
by the supernova (Burrows et al. 1995). The inner ring is believed to
represent an equatorial density enhancement in the circumstellar gas at the
interface between a dense wind emitted from an earlier red-giant phase of
the supernova progenitor star, Sk$-69^{\circ}$202, and a faster wind emitted
by this star in more recent times (Crotts, Kunkel \& Heathcote 1995; Plait
et al. 1995). In the past few years, there has been increasing evidence for
interaction of the expanding ejecta or the associated shocks with the
equatorial ring, with small regions of enhanced H$\alpha$ emission, or
`hotspots' appearing just inside the ring. The first of these was detected
by Pun et al. (1997), but studies of ground-based and {\em Hubble Space
Telescope}\ ({\em HST}) data by Lawrence et al. (2000) show that this spot
was detectable as far back as March 1995 (day 2933 since the
supernova\footnote{Day number = MJD $-46849.3$}). Lawrence et al. also
present evidence for up to eight additional regions of enhanced emission
from about day 4300 (December 1998). The most prominent are the original
spot at position angle $29^{\circ}$ and a group of spots just south of east
between position angles of $90^{\circ}$ and $140^{\circ}$.


At radio wavelengths, the initial outburst was very short-lived (Turtle et
al. 1987) compared to other radio supernovae (e.g. Weiler et al. 1998). This
prompt outburst is attributed to shock acceleration of synchrotron-emitting
electrons in the stellar wind close to the star at the time of the explosion
(Storey \& Manchester 1987; Chevalier \& Fransson 1987). After about three
years, in mid-1990, radio emission was again detected from the supernova,
with the Molonglo Observatory Synthesis Telescope (MOST) at 843 MHz (Ball et
al. 1995) and with the Australia Telescope Compact Array (ATCA) at 1.4, 2.4,
4.8 and 8.6 GHz (Staveley-Smith et al. 1992; Gaensler et al. 1997). This
emission has increased more-or-less monotonically since its first
detection. The spectral index over the observed frequency range has remained
close to $-0.9$, indicating optically thin synchrotron emission. X-ray
emission was observed to turn on at about the same time as the second phase
of radio emission and also has increased in intensity since then
(Gorenstein, Hughes \& Tucker 1994; Hasinger, Aschenbach \& Tr\"umper
1996). Recent observations with the {\em Chandra X-ray Observatory}\ have
resolved the X-ray emission into an approximately circular shell (Burrows et
al. 2000). This second phase of increasing emission is attributed to the
interaction of shocks driven by the ejecta with circumstellar material and
is distinct from the interaction with a radially decreasing stellar wind
which characterises radio supernovae. It therefore signifies the birth of
the {\it remnant} of SN 1987A -- the first observation of the birth of a
supernova remnant. We use the name SNR 1987A for the remnant.


By late-1992, SNR 1987A was sufficiently strong to image at 9~GHz using the
ATCA (Staveley-Smith et al. 1993a). This image, which exploited
super-resolution to resolve the source, showed that the remnant was roughly
circular with a diameter of about 0\farcs8, fitting inside the optical
equatorial ring (Reynolds et al. 1995), and with bright lobes to the east
and west, suggesting an annular structure with an axis similar to that of
the optical emission. The eastern lobe was about 20\% brighter than the
western lobe. To follow the evolution of this structure, the remnant has
been imaged at roughly yearly intervals since 1992.  

Despite the increasing brightness, the size of SNR 1987A is increasing
only slowly. Gaensler et al. (1997) fitted a model consisting of a
thin spherical shell to the $uv$-plane data and showed that, between
1992 and 1995, the average expansion velocity of the remnant was only
$2800 \pm 400$ km s$^{-1}$. In contrast, the average expansion speed
between 1987 and 1991 was about 35,000 km s$^{-1}$ (Jauncey et
al. 1988). Comparison of four images obtained between 1992 and 1995
showed that the brightness of the lobes increased relative to that of
the spherical shell and that the asymmetry between the east and west
lobes increased markedly over this period (Gaensler et al. 1997). By
1995, the peak brightness of the eastern lobe was 1.8 times that of
the western lobe.

The only other young supernova which has been imaged with high
resolution at radio wavelengths is SN 1993J in M81. Frequent observations
using Very Long Baseline Interferometry (VLBI) techniques (e.g. van
Dyk et al. 1994, Marcaide et al. 1997, Bartel et al. 2000) have shown
that the radio emission is in the form of an expanding shell, the
outline of which is nearly circular, but rather lumpy. Unlike SNR
1987A, SN 1993J remains in the radio supernova stage with decreasing
flux density, making further imaging difficult. Recent observations of
SN 1980K (Montes et al. 1998) and SN 1979C (Montes et al. 2000) have
shown variations in the rate of decline of flux density or, in the
case of SN 1979C, possibly small increases, indicating variations in
the density of the circumstellar medium, but, like SN 1993J, these
objects are best considered to be still in their radio supernova
stage.

The slow expansion velocity of SNR 1987A suggests that the expanding shock
and the leading ejecta have encountered a significant density enhancement,
greatly reducing their velocity (Chevalier 1992; Duffy, Ball \& Kirk 1995;
Chevalier \& Dwarkadas 1995). These models assume spherical symmetry and
hence do not account for the increasing asymmetry of the
source. Furthermore, they do not predict the continuing increase in
brightness of the remnant as observed by Ball et al. (1995) and discussed
further below. An alternative explanation for the slow apparent expansion
velocity is that the shock excites slowly moving clumps of circumstellar
material and then moves on. Ball \& Kirk (1992) modelled the emission
observed up to day 1800 by shock heating of two clumps and obtained a good
fit to the data up to that time.

As discussed by Gaensler et al. (1997), it is not easy to account for the
observed asymmetry of SNR 1987A. Models involving the annular structure of
the circumstellar material (e.g.  Chevalier \& Dwarkadas 1995) have
difficulty accounting for the degree of enhancement in the lobes and the
east-west asymmetry. Similar difficulties are encountered in trying to
explain the optical hotspots. Lawrence et al. (2000) conclude that
``fingers or jets'' in the distribution of ejecta from the supernova is the
most plausible explanation.


In a recent publication, Ball et al. (2001) have analysed the 843 MHz flux
densities observed with MOST to 2000 May (day 4820). They find a transition from a
declining rate of increase observed from about mid-1991 (day $\sim 1600$) to
early-1995 (day $\sim 2900$), to a larger and constant rate of increase,
$62.7 \pm 0.5 \;\mu$Jy day$^{-1}$, since then. This change in slope occurred
at about the same time as the appearance of the first optical hotspot
(Lawrence et al. 2000), suggesting a possible connection.

In this paper, we extend the ATCA observational data base from 1995 to
February 2001 (day 5100) and discuss these results in conjunction with those
of Ball et al. (2001). The evolution of the radio flux densities is
discussed in Section 2. The sequence of images is extended to late-2000 and
discussed in relation to recent optical and X-ray imaging in Section
3. Future prospects are canvassed in Section 4.


\section{Evolution of Radio Flux Densities and Spectral Index}
Flux density monitoring observations of SNR 1987A are made using the
ATCA at 4 -- 6 week intervals using one of the 6-km array
configurations. Observations are made simultaneously at two
frequencies, either 1380 and 2496 MHz (2368 MHz before mid-1997) or
4790 and 8640 MHz. A 128-MHz bandwidth is observed at all
frequencies. The two frequency pairs are observed alternately, with 20
min on SNR 1987A and 3 min on phase calibrators before and after the
SNR observation, with a total observation time typically of 12
hours. All observations are made with a J2000 pointing and phase
centre of R.A. $05^{\rm h}\;35^{\rm m}\;27\fs90$,
Dec. $-69^{\circ}\;16'\;21\farcs6$, approximately $10''$ south of
the SN 1987A position (Reynolds et al. 1995). The phase calibrators
are PKS~B0530--727, PKS~B0407--658 and (at 4790/8640 MHz)
PKS~B0454-810. Flux calibration is relative to PKS~B1934-638, assumed
to have flux densities of 14.95, 11.14, 5.83 and 2.84 Jy at the four
frequencies, respectively.

Data are first checked for obvious interference or telescope problems,
flagged if necessary and then are processed using {\sc miriad}\footnote{See
http://www.atnf.csiro.au/computing/software/miriad/} scripts to ensure
consistency. Visibility data are calibrated in both phase and amplitude and
images formed using baselines longer than 3 k$\lambda$. These images are
cleaned and sources above a threshold identified. A table of source
positions and integrated and peak flux densities is output for each
frequency. At least at the two higher frequencies, SNR 1987A is resolved and
so integrated flux densities are quoted. An unresolved source, J0536-6919,
is present on all images with an observed flux density ranging between 80
mJy at 1.4 GHz and 6 mJy at 8.6 GHz and serves as a check on the system
calibration. Its J2000 position determined from the 4.8 GHz observations
since day 4000 is R.A. $05^{\rm h}\;36^{\rm m}\;04\fs789 \pm 0\fs005$,
Dec. $-69^{\circ}\;18'\;44\farcs81 \pm 0\farcs02$. At 8.6 GHz it lies at
about 1.5 primary beam radii from the pointing centre, so it's flux density
measurements at this frequency are not very reliable. Measured flux
densities for SNR 1987A are given in Appendix 1.  

Figure~\ref{fg:fluxes} shows the measured flux densities of SNR 1987A at the
four ATCA frequencies. Estimated uncertainties represent a combination of random
noise and scale errors resulting from errors in the calibration. Except at
8.6 GHz, the scale errors are estimated from the scatter in the measured
flux densities of J0536-6918 since day 4000. At 8.6 GHz, J0536-6918 is outside
the half-power radius of the primary beam, and scale errors are taken to be
1.25 times the 4790 MHz scale errors. At the lower frequencies and at later
times, the errors in the flux density estimations are dominated by the scale
errors.  Since, except at 8.6 GHz, the flux density of J0536-6918 is
comparable to that of SNR 1987A and since there is no evidence for
systematic changes in the flux density of J0536-6918, these scale errors can
be reduced by dividing the flux densities by the normalised flux density of
J0536-6918 from the same observation. Flux densities from day 3000 scaled in
this way for 1.4, 2.4 and 4.8 GHz are shown in Figure~\ref{fg:scflux}.

\begin{figure}[t]
\begin{center}
\psfig{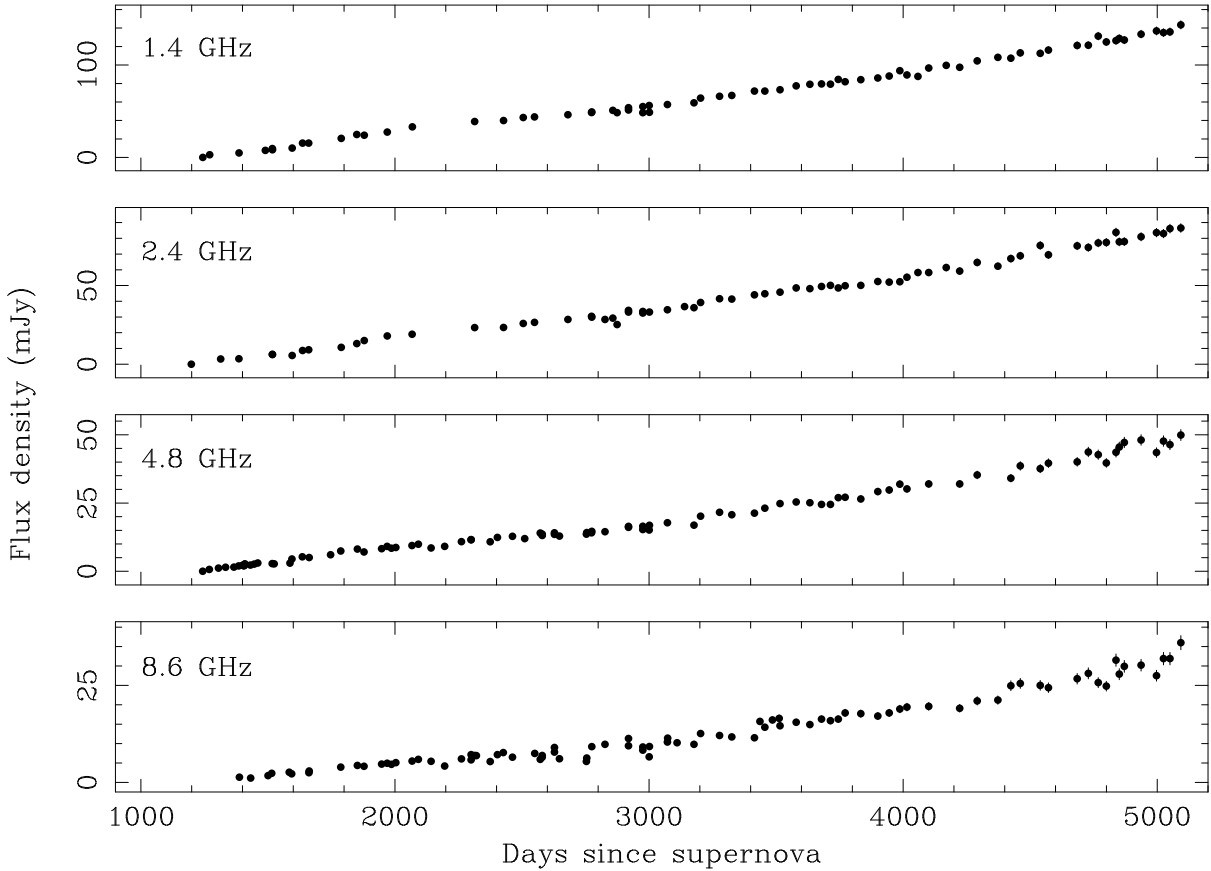}
\caption{Flux densities of SNR 1987A from 1990 August to
2001 February observed with the Australia Telescope Compact Array.}
\label{fg:fluxes}            
\end{center}
\end{figure}

\begin{figure}[t]
\begin{center}
\psfig{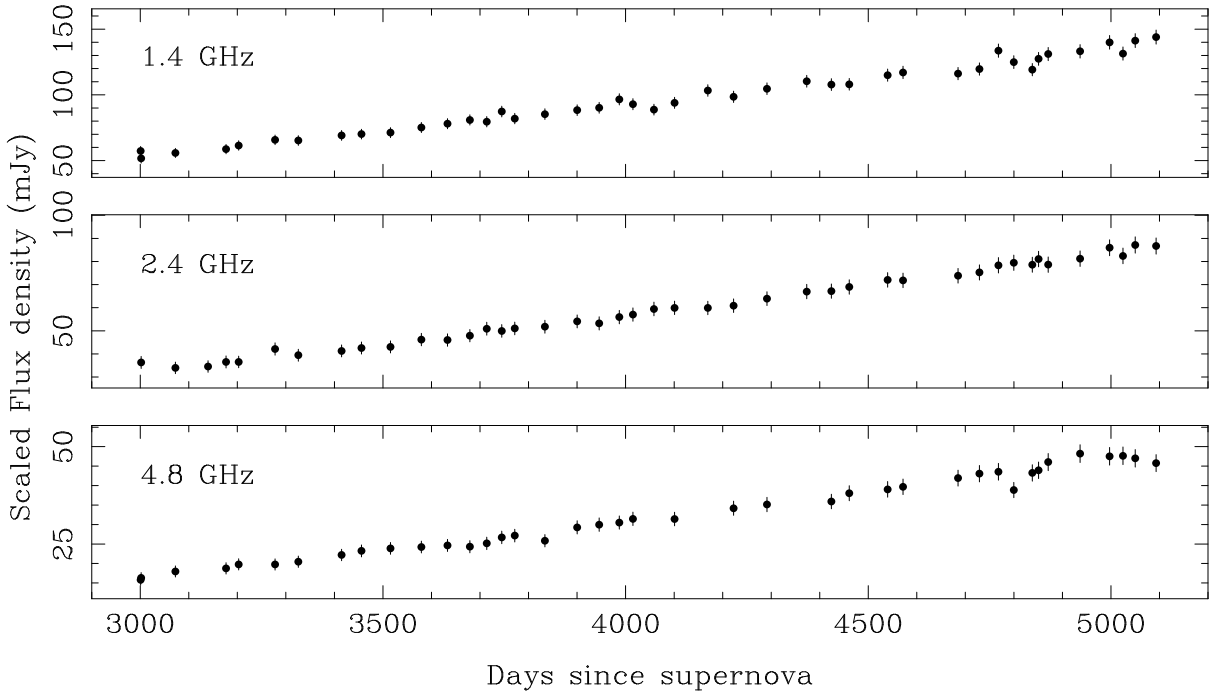}
\caption{Flux densities of SNR 1987A at 1.4, 2.4 and 4.8 GHz from 1995 May to
2001 February, scaled by the flux density of J0536-6918 obtained from the same
image. This scaling removes most of the effects of mis-calibration.}
\label{fg:scflux}            
\end{center}
\end{figure}

\begin{figure}[t]
\begin{center}
\psfig{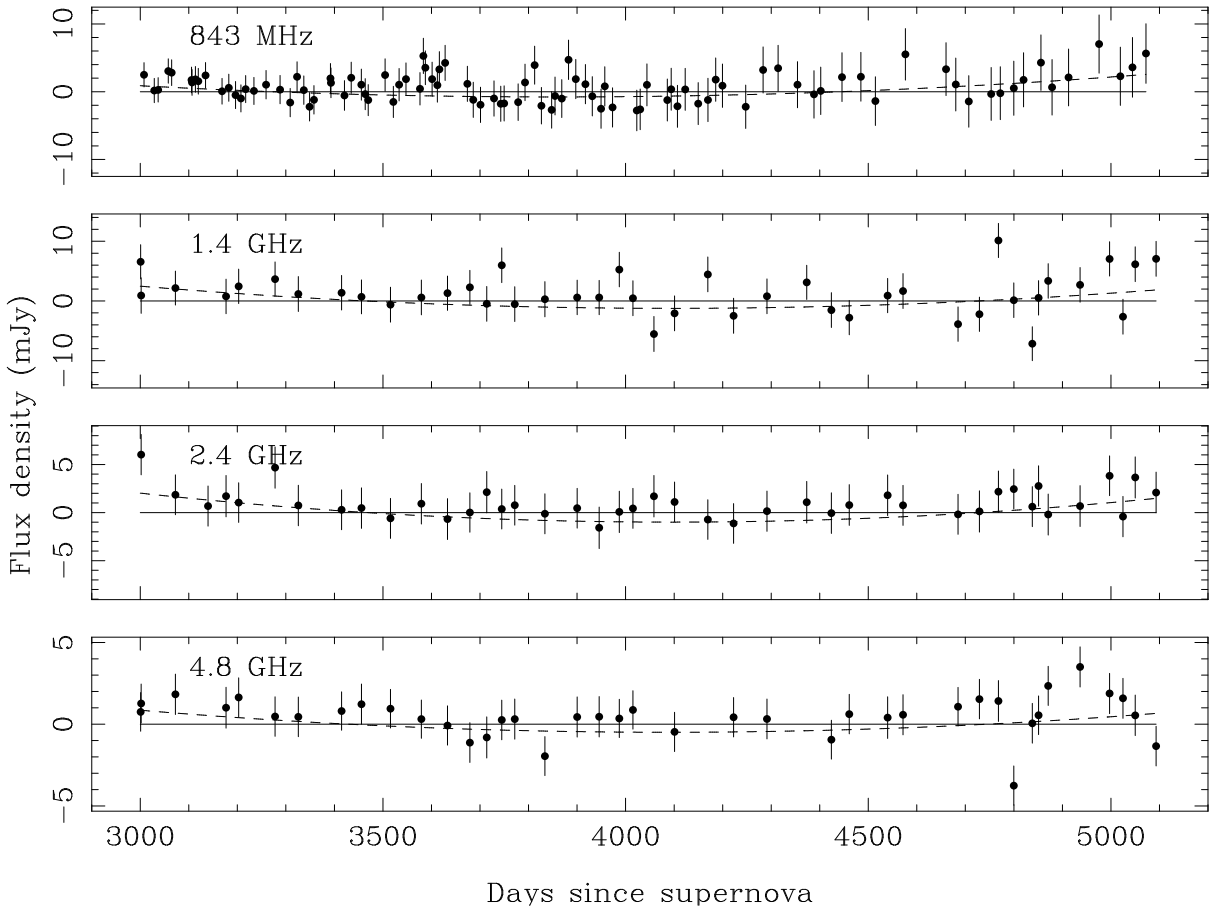}
\caption{Residual flux densities of SNR 1987A from 1990 August to 2001
February after fitting a second-order polynomial to the data and subtracting
the constant and linear terms. The upper panel shows data from the MOST
(Ball et al., 2001) and the lower three panels show ATCA data scaled by the
J0536-6918 flux density. In each case the second-order term is shown by a
dashed line. }
\label{fg:fluxvar}            
\end{center}
\end{figure}

These plots show that the continued increase in flux density observed
at 843 MHz by Ball et al. (2001) is also observed at higher
frequencies. The increase in slope observed at about day 2900 in the
MOST data is also seen in the ATCA data (Fig.~\ref{fg:fluxes}). Ball
et al. (2001) stated that the MOST data after day 3000 were well
fitted by a linear trend. However, we believe that there is
significant evidence for an long-term increase in slope after day 3000
in both the MOST data set and the ATCA data sets, i.e., the rate of
increase in flux density is increasing.  Evidence for this is shown in
Fig.~\ref{fg:fluxvar} which shows residual flux densities after
fitting a second-order polynomial to the MOST and scaled ATCA flux
densities from day 3000 and subtracting the linear component. These
plots all show a systematic trend with the residual flux densities
being, on average, negative at central times and positive at both
early and late times. Fig.~\ref{fg:fluxvar} also shows the
second-order term of the fit which in all cases is positive and of
about 3-$\sigma$ significance. These data sets are essentially
independent so there can be little doubt about the significance of the
effect.

To further quantify this effect, straight lines were fitted to the ATCA data
sets from day 3000, both for the unscaled data and for the data (except at
8.6 GHz) scaled by the flux density of J0536-6918. Results of this fitting
are given in Table~\ref{tb:grad}. The second and third columns give the
gradient and $x$-intercept of the lines fitted to all data between day 3000
and day 5100 (cf. Fig.~\ref{fg:fluxvar}). The point of interest here is that
there is a systematic increase of the date of intercept with
frequency. Especially for the higher frequencies, the intercept is well
after the date of first detection of the SNR. This shows that
the higher frequencies have a higher relative rate of flux density increase
and that the present rate of increase is much higher than that at early
times. Note that this effect is seen in both the scaled and unscaled data,
so it is not an artifact of the scaling procedure.

As a further check, the data sets were split into two halves, from day 3000
to day 4050 and from day 4050 to day 5100, and straight lines fitted to both
halves. Results are tabulated in columns 4 -- 7 of Table~\ref{tb:grad}. The
values of reduced $\chi^2$ for the combined fit were computed by treating
the two lines as a single model describing the complete data set from day
3000 onward.  These show that, independently at all frequencies, the change
in gradient is significant at about the 2-$\sigma$ level, and that this
gradient increase is present in the MOST data and in both the scaled and the
unscaled ATCA data.

\begin{table}[htb]
\begin{center}
\caption{Flux density gradients since day 3000}
\label{tb:grad}
\begin{tabular}{ccccccc} \hline
Frequency & Gradient & Intercept & Grad. ($<4050$)  & Grad. ($>4050$) & Grad. change& 
Red. $\chi^2$\\
 (GHz) & ($\mu$Jy/day) & (day) &  ($\mu$Jy/day)  & ($\mu$Jy/day) & ($\mu$Jy/day)&  \\ \hline
\multicolumn{7}{c}{Unscaled} \\ 
0.843 & $63.1\pm 0.6$ & $1640\pm 20$ & $61.4\pm 1.0$  &  $66.9\pm 2.1$ & $5.5\pm 2.3$ & 0.5\\
1.4 & $40.9\pm 0.8$ & $1750\pm 50$ & $37.7\pm 1.7$  &  $46.6\pm 2.4$ & $8.8\pm 2.9$ & 0.7\\
2.4 & $25.2\pm 0.5$ & $1780\pm 50$ & $20.7\pm 1.2$  &  $28.0\pm 1.6$ & $7.3\pm 1.9$ & 0.7\\
4.8 & $15.2\pm 0.4$ & $2010\pm 55$ & $14.2\pm 0.6$  &  $17.5\pm 1.2$ & $3.2\pm 1.4$ & 1.3\\
8.6 & $10.6\pm 0.4$ & $2160\pm 65$ & $10.3\pm 0.8$  & $ 13.3\pm 1.7$ & $3.0\pm 1.9$ & 0.8 \\
\multicolumn{7}{c}{Scaled by J0536-6918} \\ 
1.4 & $41.2\pm 0.7$ & $1770\pm 45$ & $39.8\pm 2.0$  &  $46.3\pm 2.0$ & $6.4\pm 2.8$ & 1.3\\
2.4 & $26.0\pm 0.5$ & $1840\pm 45$ & $22.8\pm 1.5$  &  $27.8\pm 1.4$ & $5.0\pm 2.0$ & 0.4\\
4.8 & $15.3\pm 0.3$ & $2020\pm 45$ & $13.9\pm 0.8$  &  $15.5\pm 1.0$ & $2.5\pm 1.3$ & 1.0\\
\hline
\end{tabular}
\end{center}
\end{table}

Apparently significant short-term variations are seen, especially at 4.8
GHz, with a timescales of order 100 days. The clearest example is near the
end of the 4.8 GHz data set and is best seen in
Figure~\ref{fg:fluxvar}. This fluctuation is not obvious at lower
frequencies, but could be masked by the low signal-to-noise ratio. If real,
these short-term fluctuations imply significant interactions on a scale of
0.03 pc (0\farcs1) or less, and furthermore, many such interactions.

\begin{figure}[h]
\begin{center}
\psfig{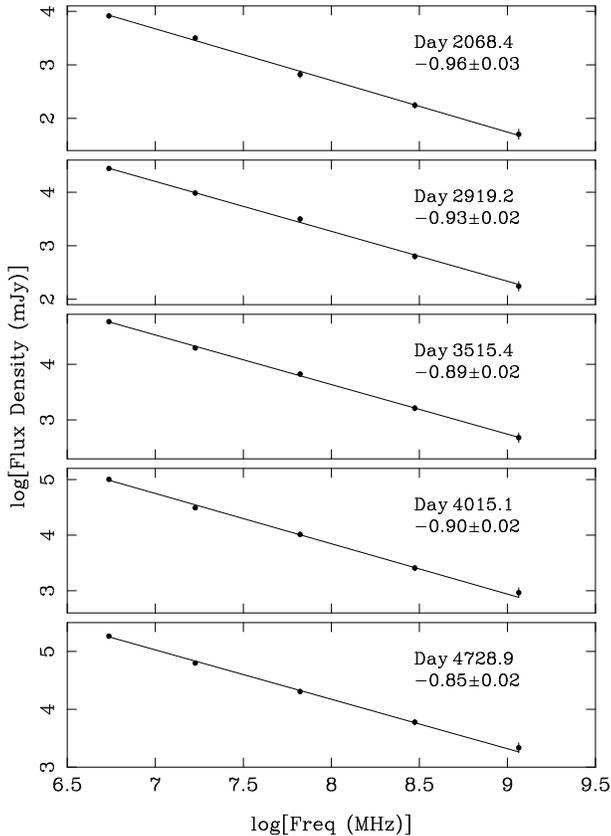}
\caption{The radio spectrum of SNR 1987A at five different epochs}
\label{fg:spec_ind}
\end{center}
\end{figure}

Figure~\ref{fg:spec_ind} shows MOST flux densities from Ball et
al. (2001) and from the ATCA at 1.4, 2.4, 4.8 and 8.6 GHz at five epochs
spread through the data set. Values of the spectral index $\alpha$, where
$S=\nu^{\alpha}$ and $\nu$ is the frequency, found by linear regression, are
given on each plot. Quoted errors are 1$\sigma$.

To test for systematic curvature in the spectrum, the spectral indices were
calculated using the three lowest frequencies, $\alpha_1$ and separately
using the three highest frequencies, $\alpha_2$. If there is systematic
curvature in the spectrum, then the difference $\alpha_1 - \alpha_2$ should
be significant, and roughly constant. In fact, the difference is typically
small and of either sign, showing that there is no systematic
curvature. The mean difference $\alpha_1 - \alpha_2$ across the entire data set
is $-0.035$ compared to the rms fluctuation of $0.187$.

\begin{figure}[h]
\begin{center}
\psfig{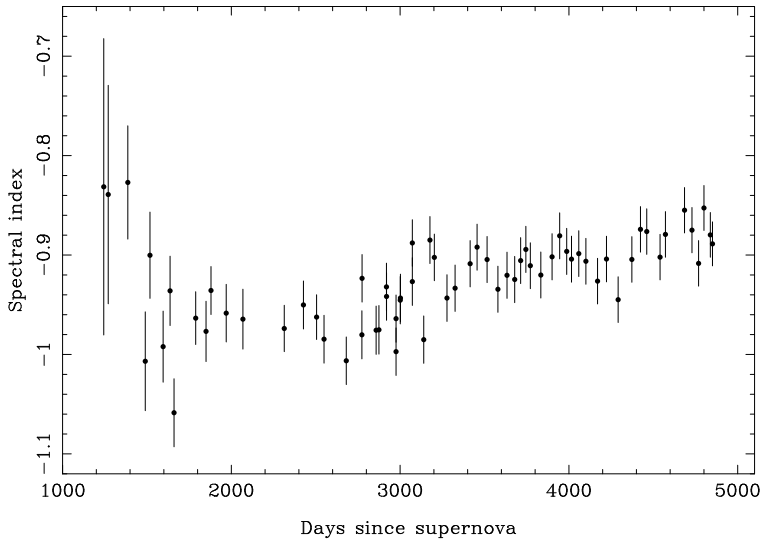}
\caption{Spectral indices of the radio emission from SNR 1987A as a function
of time obtained by combining the MOST 843 MHz data of Ball et al. (2001)
with ATCA data at 1.4, 2.4, 4.8 and 8.6 GHz.}
\label{fg:si_time}            
\end{center}
\end{figure}

The observation that the relative flux density gradient is both greater and
increasing more rapidly at the higher frequencies (Table~\ref{tb:grad}) implies
a systematic change in the spectral index.  Figure~\ref{fg:si_time} shows
the computed spectral indices from 843 MHz to 8.4 GHz as a function of
time. Because of the uneven and non-simultaneous sampling at the different
frequencies, spline curves were fitted to all data sets except that at 1.4
GHz. Spectral indices were then computed by interpolating values at 843 MHz,
2.4, 4.8 and 8.6 GHz to times when the 1.4 GHz flux density was measured,
and fitting power-law spectra to flux densities. The interpolation was
unreliable near the start of the data set for 2.4 and 8.6 GHz because of
sparse data, so only three points were fitted there.

Apart from a few high points between days 1200 and 1400 near the start of
the data set (which have large error bars), there is a more-or-less steady
increase in spectral index, corresponding to a flattening of the radio
spectrum, throughout the whole data set. It is possible that the
variation consists of step changes, with the most obvious step times being
around day 3000 and day 4700. At early times, the mean spectral index was
about $-0.97$, but in the last year (2000) it was $-0.88$. 

The radio spectra are remarkably close to power law and show no sign of
either positive curvature, as would be expected if either free-free
absorption or synchrotron self-absorption were important, or negative
curvature as is predicted by diffusive shock acceleration theories
(e.g. Reynolds \& Ellison 1992). The radio spectral index of
about $-0.9$ is considerably steeper than the canonical $-0.5$ expected for
synchrotron emission from relativistic electrons accelerated in strong
shocks, for which the spectral index $\alpha = (1-s)/2$ and the electron
energy index $s = (r+2)/(r-2)$, where $r$ is the compression ratio in the
shock (Jones \& Ellison 1991). For high Mach-number shocks in a
monatomic non-relativistic gas, $r=4.0$, giving $s=2.0$ and $\alpha =
-0.5$. 


A steeper radio spectrum implies a steeper electron energy distribution and
a smaller compression ratio in the shock; $s \sim 2.8$ and $r \sim 2.7$ for
$\alpha = -0.9$. Monte-Carlo modelling of non-linear diffusive shock
acceleration, including the dynamical effects of the accelerated particles
(e.g. Baring et al. 1999), suggest that young and hence fast
shocks have a {\it larger} compression ratio than older shocks, resulting in
a flatter synchrotron spectrum. Duffy, Ball \& Kirk (1995)
included the effect of accelerated ions on the shock structure and were able
to model the observed spectrum with the shock expanding into a terminated
stellar wind structure. However, this model predicts a declining brightness
and a steepening spectrum for the radio emission, both of which are contrary
to observation. 

The observed spectral index is also steeper than those for typical
supernova remnants; no SNR in the Green (2000) catalogue has a
spectral index definitely steeper than $-0.8$ and the mean spectral index
for shell-type remnants is $-0.51$. It is interesting to speculate that the
current flattening of the spectrum represents an evolution toward the
typical index of $-0.5$. At the current rate, it will only take about 50 years
to reach $-0.5$.


The clump interaction model of Ball \& Kirk (1992) predicts a
declining flux density at later times. However, this model was based
on only two clumps. The model could be extended to invoke interaction
with an increasing number of clumps, maybe tens or hundreds at the
present time. A more accurate description would involve a statistical
hierarchy of effective clump sizes. This could account for both the
gradient increase and the short-term fluctuations in the observed flux
density. Both radio and optical evidence (e.g. Spyromilio, Stathakis
\& Meurer 1993, van Dyk et al. 1994, Spyromilio 1994) point to a
clumpy circumstellar medium around other supernovae.

\section{9~GHz Imaging Observations of SNR~1987A}

\subsection{Resolved Radio Images of SNR~1987A}

In previous papers we have shown that the ATCA's diffraction-limited
resolution at 9~GHz of 0\farcs9 is sufficient to
resolve the radio emission from SNR~1987A (Staveley-Smith et
al. 1993b) and that super-resolution techniques can be
used to improve the resolution to $\sim0\farcs5$. At
this resolution, the radio emission forms a limb-brightened shell, with
brightness enhancements on the eastern and western sides (Staveley-Smith et
al.  1993a; Briggs 1994; Gaensler et al. 1997).


We have continued to make regular observations of SNR~1987A at 9~GHz;
observing parameters for all imaging observations are given in
Table~\ref{tab_9ghz}.  We have analysed these data in the same manner as
described by Gaensler et al. (1997), forming an image at each epoch and
deconvolving the resulting image using a maximum entropy algorithm.  From
1996 onwards, the source has been of sufficient signal-to-noise that phase
self-calibration can be successfully applied to the data, resulting in a
significant improvement in the accuracy of the complex gains for each
antenna.

\begin{table}[htb]
\begin{center}
\caption{9~GHz ATCA observations of SNR~1987A used for imaging. The
radius listed is that obtained by fitting a thin spherical shell
to each $u-v$ data-set (see text).}
\label{tab_9ghz}
\begin{tabular}{cccccccc} \hline
Mean Epoch & Observing &   Day & Array & Frequencies   & Time on &
Radius \\
 &   Date   &   Number & & (MHz)  & Source (hr) &   ($''$) \\  \hline
1992.9 & 1992 Oct 21  &   2068  & 6C & 8640,8900 &  15  &   0.66(2) \\
       & 1993 Jan 04  &   2142  & 6A & 8640,8900 &  13  &      0.66(2)   \\
       & 1993 Jan 05  &   2143  & 6A & 8640,8900 &   5  &      0.62(2)     \\ \hline
1993.6 & 1993 Jun 24  &   2314  & 6C & 8640,8900 &   9  & 0.62(1)  \\
       & 1993 Jul 01  &   2321  & 6C & 8640,8900 &  10  &     0.68(1)  \\
       & 1993 Oct 15  &   2426  & 6A & 8640,9024 &  18  &    0.69(1) \\ \hline
1994.4 & 1994 Feb 16  &   2550  & 6B & 8640,9024 &   9  & 0.69(2) \\
       & 1994 Jun 27  &   2683  & 6C & 8640,9024 &  21  &   0.659(7)  \\
       & 1994 Jul 01  &   2687  & 6A & 8640,9024 &  10  &  0.659(9) \\ \hline
1995.7 & 1995 Jul 24  &   3074  & 6C & 8640,9024 &   7  & 0.687(8)\\
       & 1995 Aug 29  &   3111  & 6D & 8896,9152 &   7  &   0.69(2) \\
       & 1995 Nov 06  &   3178  & 6A & 8640,9024 &   9  &  0.685(6) \\ \hline
1996.7 & 1996 Jul 21  &   3437  & 6C & 8640,9024 &  14  &  0.688(4) \\
       & 1996 Sep 08  &   3486  & 6B & 8640,9024 &  13  &     0.684(4) \\
       & 1996 Oct 05  &   3512  & 6A & 8896,9152 &   8  &     0.692(6) \\ \hline
1998.0 & 1997 Nov 11  &   3914  & 6C & 8512,8896 &  19  &  0.715(5) \\
       & 1998 Feb 18  &   4013  & 6A & 8896,9152 &  15  &     0.733(5) \\
       & 1998 Feb 21  &   4016  & 6B & 9024,8512 &   7  &     0.735(4) \\ \hline
1998.9 & 1998 Sep 13  &   4220  & 6A & 8896,9152 &  12  &    0.721(3) \\
       & 1998 Oct 31  &   4268  & 6D & 9024,8512 &  11  &     0.729(5) \\
       & 1999 Feb 12  &   4372  & 6C & 8512,8896 &  10  &     0.737(3) \\ \hline
1999.7 & 1999 Sep 05  &   4578  & 6D & 9152,8768 &  11  &   0.754(4) \\
       & 1999 Sep 12  &   4585  & 6A & 8512,8896 &  14  &      0.736(4) \\ \hline
2000.8 & 2000 Sep 28  &   4966  & 6A & 8512,8896 &  10  & 0.756(2) \\ 
       & 2000 Nov 12  &   5011  & 6C & 8512,8896 &  11  & 0.764(3) \\ \hline
\end{tabular}
\end{center}
\end{table}

\begin{figure}[htb]
\begin{center}
\psfig{file=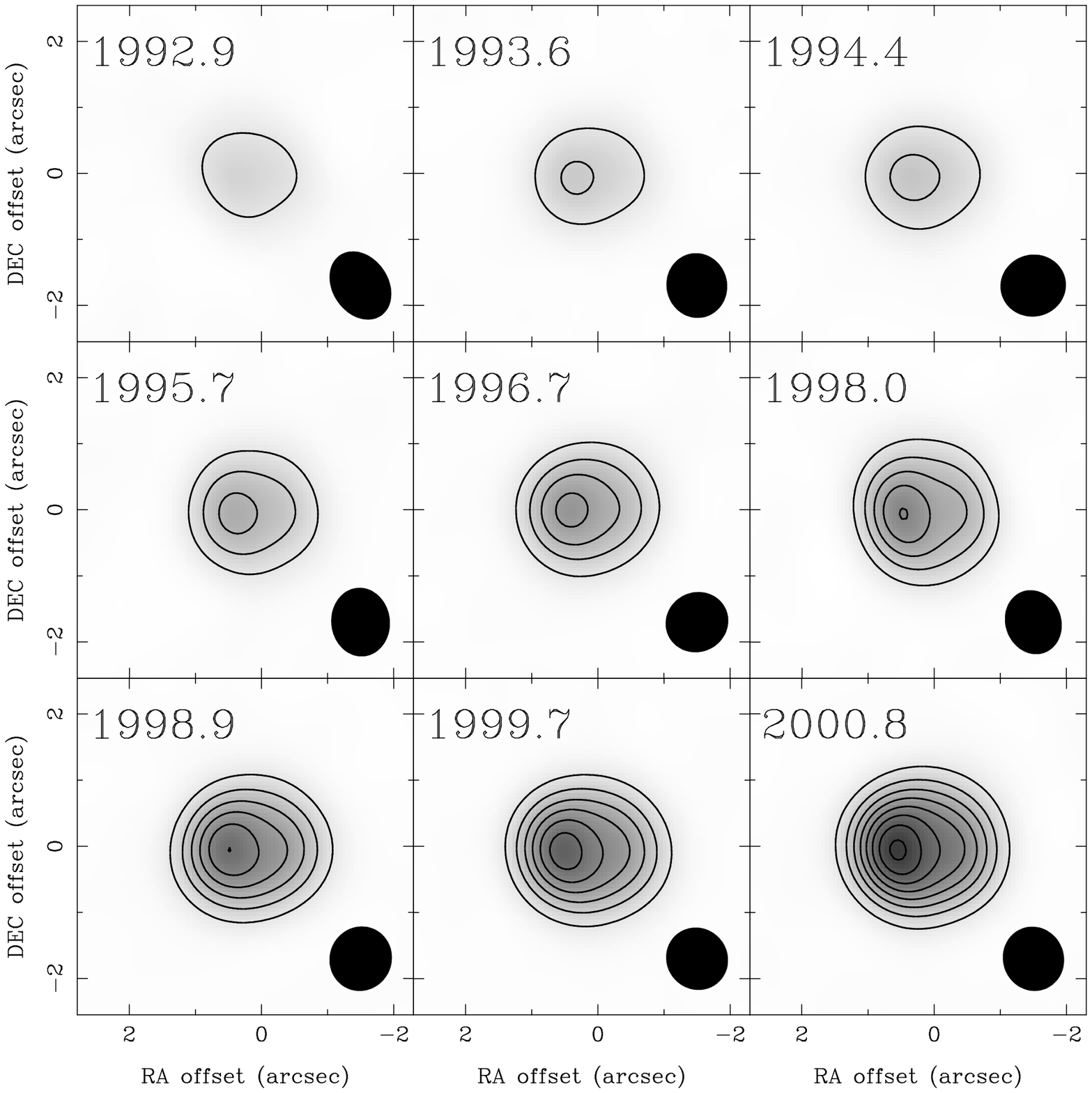,angle=270}
\caption{Diffraction-limited 9~GHz images of SNR~1987A at the nine epochs
listed in Table~\ref{tab_9ghz}. The false-colour range in each
panel is --0.1 to $+$16.0~mJy~beam$^{-1}$, while contours
are at levels of 1.5, 3.0, 4.5, $\ldots$, 13.5~mJy~beam$^{-1}$.
The FWHM dimensions of the diffraction-limited synthesised beam
for each epoch is shown at the lower-right of each panel.}
\label{fig_movie_natural}
\end{center}
\end{figure}

\begin{figure}[htb]
\begin{center}
\psfig{file=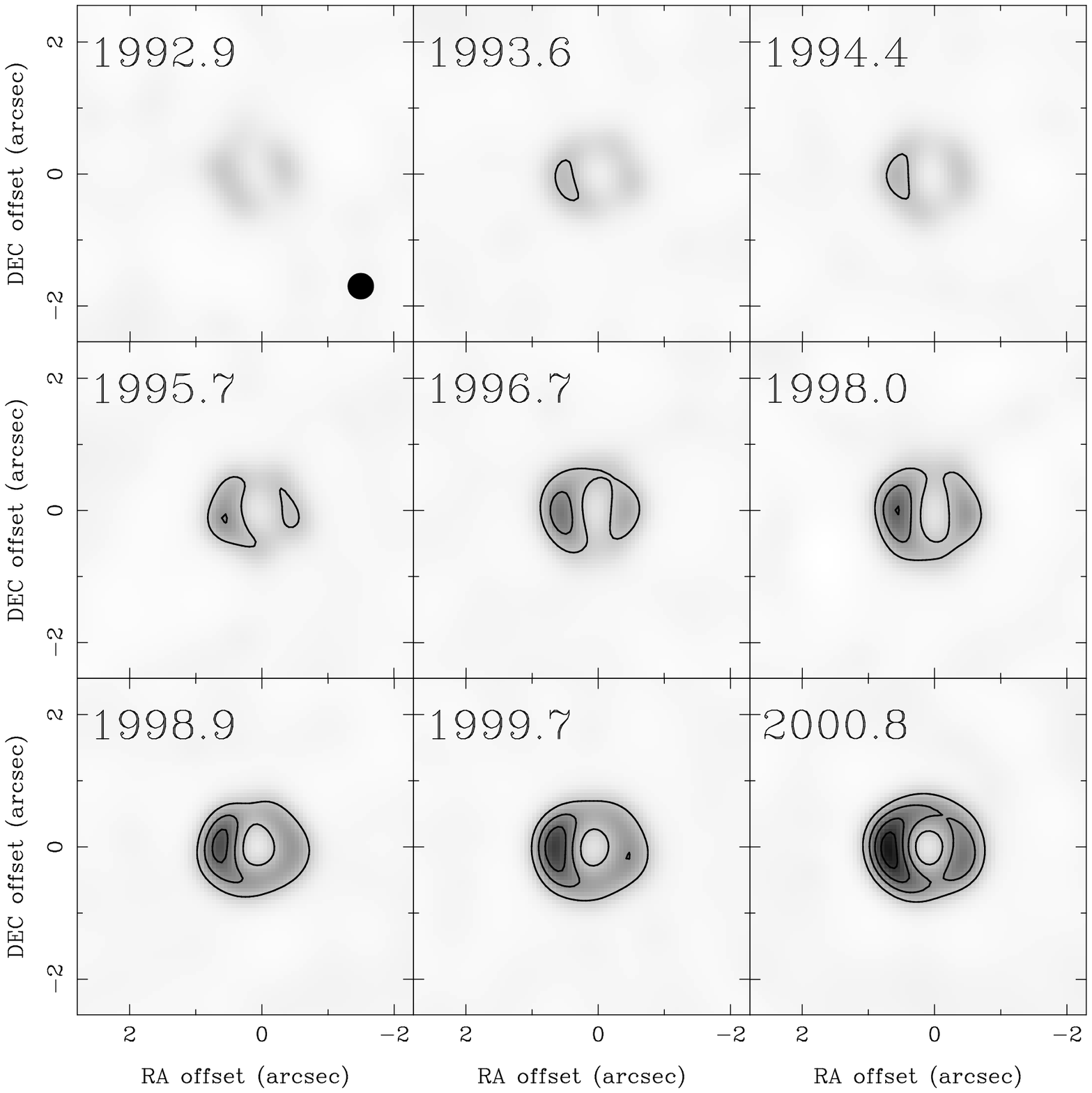,angle=270}
\caption{Super-resolved 9~GHz images of SNR~1987A at the nine epochs
listed in Table~\ref{tab_9ghz}. The false-colour range in each
panel is --0.1 to $+$5.0~mJy~beam$^{-1}$, while contours
are at levels of 0.4, 0.8, 1.2, $\ldots$, 2.8~mJy~beam$^{-1}$.
The FWHM dimensions of the super-resolved synthesised beam is shown at
the lower-right of the first panel.}
\label{fig_movie_super}
\end{center}
\end{figure}

The resulting series of images are shown in Figures~\ref{fig_movie_natural}
and \ref{fig_movie_super} under the conditions of diffraction-limited
resolution and super-resolution, respectively. In these figures and
Figure~\ref{fig_models}, the R.A. and Dec. offsets are with respect to J2000
R.A. $05^{\rm h}\;35^{\rm m}\;28\fs00$, Dec. $-69^{\circ}\;16'\;11\farcs1$
The diffraction-limited images indicate that the source is clearly extended,
primarily in the east-west direction, and continues to brighten.  In the
super-resolved sequence of images, it can be seen that the shell-like
morphology reported by Staveley-Smith et al. (1993a) and by Gaensler et
al. (1997) is maintained throughout, with two bright regions on the east and
west sides of the rim.

\subsection{Model Fits to the Radio Morphology}

Gaensler et al. (1997)
showed that the size of SNR~1987A could be quantified at each
epoch by approximating the morphology of SNR~1987A by a
thin spherical shell of arbitrary position, flux and radius.
The best-fit parameters are found by computing
the Fourier-transform of this shell, subtracting
this transform from the $u-v$ data, and then adjusting
the properties of the model until the corresponding
$\chi^2$ parameter is minimised (see also Staveley-Smith
et al. 1993b). Using such
an approach, Gaensler et al. (1997) were able to show
that the radius of the supernova remnant increased
from 0\farcs65 at epoch 1992.9 to 0\farcs68
at epoch 1995.7, corresponding to a (surprisingly low)
mean expansion velocity of $2800\pm400$~\kms\ (assuming
a distance to the supernova of 50~kpc).

\begin{figure}[t]
\begin{center}
\psfig{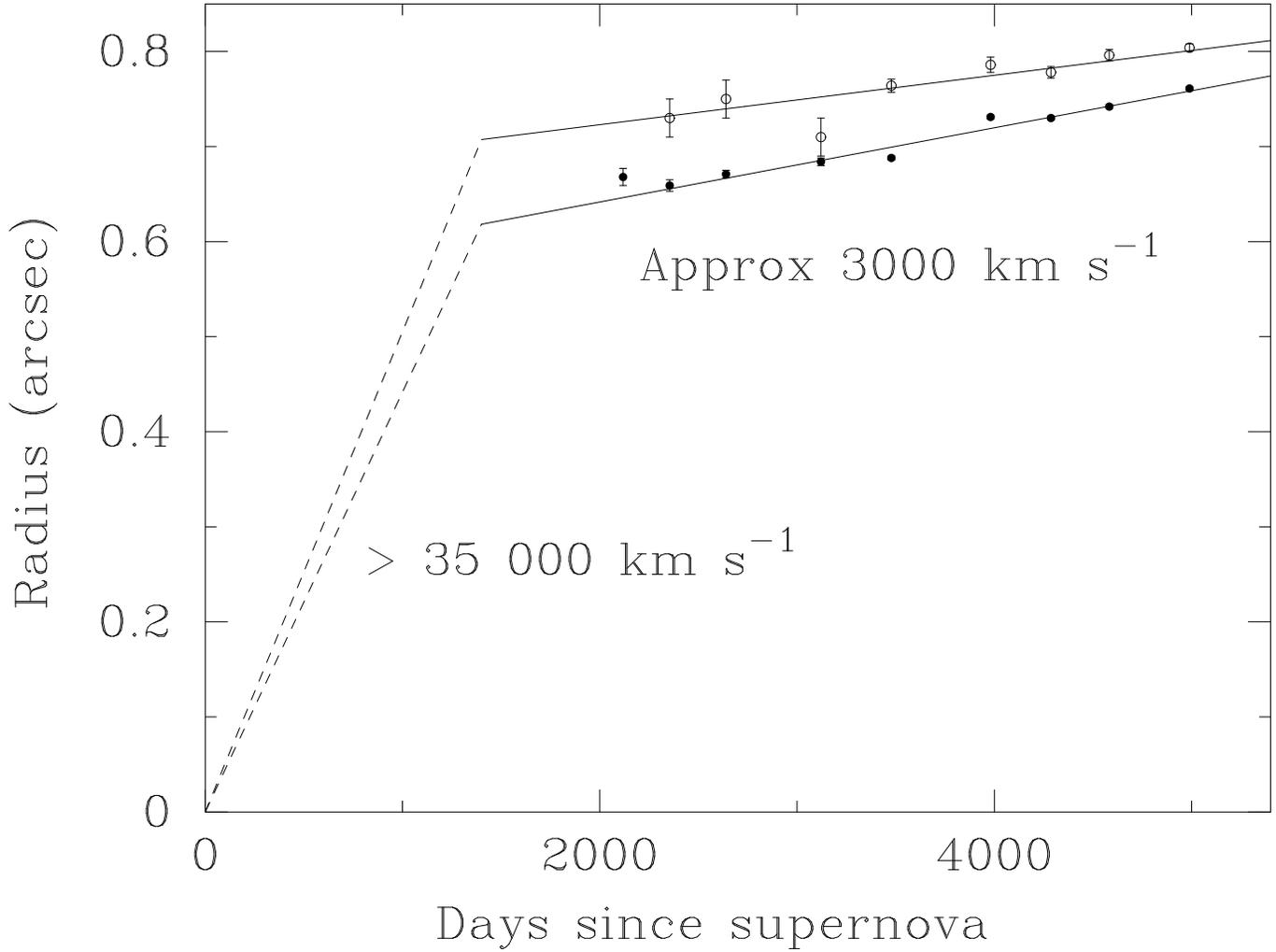}
\caption{Inferred radius of the radio remnant SNR~1987A
as a function of time. The filled points represent
radii determined by fitting a thin spherical
shell to the data, while open points correspond to
a model involving a shell plus two point sources.
The solid lines indicate the respective best-fit linear
increases, while the dashed lines are extrapolations
between the epoch of the supernova explosion and the
time when radio emission re-appeared.}
\label{fig_expand}
\end{center}
\end{figure}

Here we extend and expand on these attempts to quantify the changes in the
radio remnant. We first fit a spherical shell to all subsequent
data-sets. The resulting radii are listed in Table~\ref{tab_9ghz}, where the
uncertainty in the last quoted digit is given in parentheses. A fit to these
radii gives a linear expansion rate of $3500\pm100$~\kms. The mean radius
at each observing epoch is listed in Table~\ref{tab_params} and plotted in
Figure~\ref{fig_expand}.

\begin{figure}[t]
\begin{center}
\psfig{file=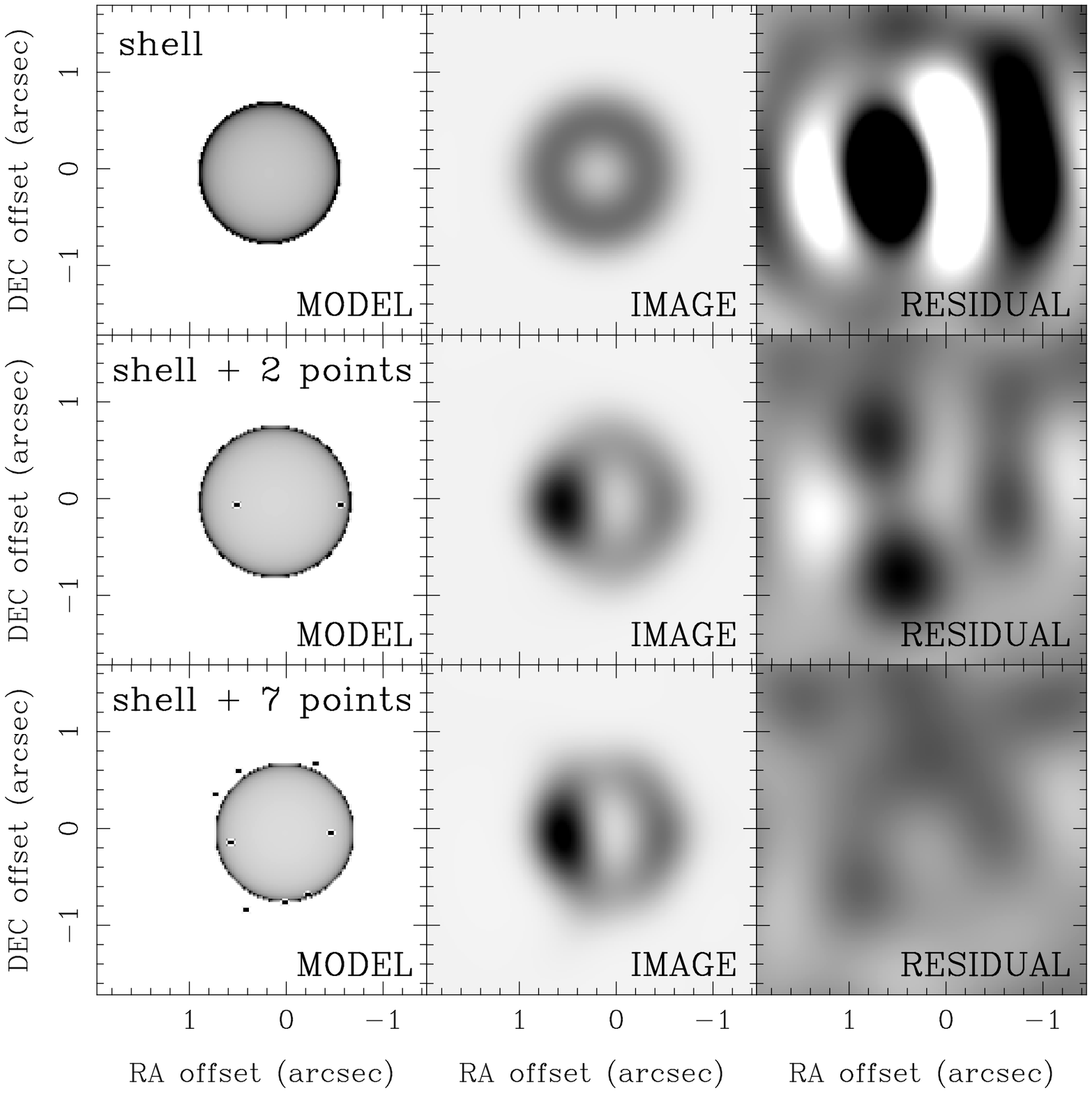,angle=270,width=18cm}
\caption{Models of the radio emission from SNR~1987A. Each row represents a
different attempt to model the radio emission from epoch 1998.0: a thin
spherical shell; a thin shell plus two hotspots; a thin shell plus seven
hotspots. The first column shows the input image for each model, using a
false-colour range of 0 to 0.01~mJy~beam$^{-1}$; the second column shows the
results of a simulated super-resolved observation of the input image, using
a grayscale range of --0.16 to +3.05~mJy~beam$^{-1}$; the third column shows
an image of the residuals remaining once the model has been subtracted from
the original data, using a false-colour range --0.2 to
+0.3~mJy~beam$^{-1}$. In the third model, the east point source within the
shell has a flux density of 3.5 mJy; the other six have flux densities in
the range 0.3 to 0.8 mJy.}
\label{fig_models}
\end{center}
\end{figure}

The signal-to-noise of the data presented by Gaensler et al. (1997) was too
low to justify more complex fits to the data. However, we are now in a
position to compare a thin spherical shell to other models. We first note
that other simple spherically-symmetric models, such as face-on rings and
gaussians, produce significant residuals, and are clearly inconsistent with
the data.  In Figure~\ref{fig_models} we show three possible fits to the
data from epoch 1998.0. In each case, we have generated an input model
(shown in the first column).  We have then simulated an observation of the
sky-distribution corresponding to this model as follows: we
Fourier-transform the model, then multiply this transform by the transfer
function of the ATCA observations from this epoch to produce a set of $u-v$
tracks identical to those of the real data.  We then image these
visibilities in the same way as for the real data, and similarly deconvolve
and super-resolve the image, to give the result shown in the second
column. We also subtract the $u-v$ data corresponding to the model from the
observed data, and then image the resulting visibilities to produce the
residual image shown in the third column.

In the first row of Figure~\ref{fig_models}, we show the results of fitting
a thin spherical shell to the 1998.0 data (cf. Gaensler et al. 1997)
--- the corresponding best-fit flux and radius are 18.4~mJy and 0\farcs73.
The resulting image successfully approximates the limb-brightened morphology
seen in Figure~\ref{fig_movie_super}, but lacks the enhancements in
brightness seen on either side of the shell. The residual image, while
having small amplitudes in an absolute sense, shows clear and systematic
differences between this simple model and the data (maximum fractional
residual of $\sim$20\%).

Given that the main difference between a thin shell model and the data seems
to be the presence of the bright lobes on either side, we next generate a
model involving a thin shell (of arbitrary flux, position and radius), and
two point-sources of emission (each of arbitrary flux and position). For the
1998.0 data set, the best-fit parameters for this model are a shell of flux
15.0~mJy and radius 0\farcs79, and point-sources of fluxes 3.1 and 0.4~mJy,
one projected near the edge of, but inside, the eastern half of the shell,
and the other similarly positioned on the western half.  Note that the
radius of the shell in this fit is 8\% larger than that obtained by fitting
to a shell alone. The resulting model, image and residual are all shown in
the second row of Figure~\ref{fig_models}.  The image corresponding to the
model now reasonably approximates the morphology seen in the real data, and
the residuals are greatly reduced (maximum fractional residual of $\sim$5\%)
when compared to the case of a shell alone.

We can further improve the fit between the model and the data by increasing
the number of point sources in the model. In the bottom row of
Figure~\ref{fig_models}, we show a fit to the 1998.0 data involving a
spherical shell plus seven point sources. The large number of free
parameters (25 in this case) makes finding an absolute minimum in $\chi^2$
very difficult --- thus the ``best fit'' we have shown was found by
trial-and-error. The shell has flux density 11.4~mJy and radius 0\farcs72
(2\% smaller than in the case of a shell alone); the point sources range in
flux density between 0.3 and 3.5~mJy, and are distributed around the
perimeter of the shell. The model image and residual image both indicate a
very good match between the model and the data (maximum fractional residual
$\sim$2\%). We emphasise, though, that we make no claims to uniqueness for
this solution.

\begin{table}[htb]
\begin{center}
\caption{Parameters from model fits to the 9 GHz images}
\label{tab_params}
{\scriptsize
\begin{tabular}{cccccccccccc} \hline
Mean & Total Flux  &  \multicolumn{2}{c}{Fit to shell alone} & 
\multicolumn{8}{c}{Fit to shell $+$ 2 points} \\
Epoch    & Density & $S_{\rm shell}$ & $r_{\rm shell}$
	 & $S_{\rm shell}$ & $r_{\rm shell}$ & $S_{\rm 1}$ & $\delta{\rm RA}_1$ & 
	 $\delta{\rm Dec}_1$ &  $S_{\rm 2}$ & $\delta{\rm RA}_2$ & $\delta{\rm Dec}_2$ \\ 
 &  (mJy) &  (mJy) & ($''$) & (mJy) & ($''$) & (mJy) & ($''$) & ($''$) & 
 (mJy) & ($''$) & ($''$) \\ \hline
1992.9 & 5.6(2)  & 5.2  & 0.668(9) & $\ldots$ & $\ldots$ & $\ldots$ &
$\ldots$ & $\ldots$ & $\ldots$ & $\ldots$ & $\ldots$ \\
1993.6 & 6.9(2)  & 6.6  & 0.659(6) & 5.1  & 0.73(2)  & 1.2 & 0.26(3) &-0.05(2) & 0.4 &-0.73(9)&  0.06(5)\\
1994.4 & 7.5(1)  & 7.8  & 0.671(4) & 6.2  & 0.75(2)  & 1.3 & 0.27(2) &-0.03(1) & 0.4 &-0.69(7)&  0.06(4)\\
1995.7 & 11.8(1) & 11.0 & 0.684(4) & 8.6  & 0.71(2)  & 1.9 & 0.35(3) &-0.03(1) & 0.5 &-0.76(8)&  0.04(5)\\
1996.7 & 15.5(1) & 15.1 & 0.688(2) & 12.4 & 0.764(7) & 2.5 & 0.32(1) & 0.03(1) & 0.4 &-0.69(6)&  0.15(3)\\
1998.0 & 18.3(1) & 18.4 & 0.731(2) & 15.0 & 0.786(8) & 3.1 & 0.34(1) &-0.02(1) & 0.4 &-0.75(7)& -0.02(5)\\
1998.9 & 21.7(1) & 21.5 & 0.730(2) & 17.2 & 0.778(6) & 3.4 & 0.40(1) &-0.02(1) & 1.0 &-0.81(3)& -0.04(2)\\
1999.7 & 24.7(1) & 24.4 & 0.742(2) & 19.7 & 0.796(6) & 3.9 & 0.41(1) &-0.02(1) & 1.0 &-0.81(3)& -0.05(2)\\
2000.8 & 30.8(1) & 30.8 & 0.761(2) & 24.9 & 0.804(4) & 4.8 & 0.46(1) &-0.01(1) & 1.3 &-0.82(2)& -0.03(2)\\ \hline
\end{tabular}}
\end{center}
\end{table}

Given the difficulty in finding a multiple-component fit to the data by
trial-and-error, and the non-uniqueness of this solution, we do not here
present multi-parameter fits to other epochs such as those shown in the
third row of Figure~\ref{fig_models}.  However, it seems clear that a shell
alone is no longer the best simple model of the data, and that a shell with
two point-sources is a significant improvement. We therefore have fitted all
epochs with these two models, with all parameters free to vary in each
case. The results of these fits are listed in Table~\ref{tab_params} and
plotted in Figure~\ref{fig_expand}. Offsets of the two point sources in
Table~\ref{tab_params} are with respect to the reference position used for
Figures 6, 7 \& 9. The data from epoch 1992.9 is of low signal-to-noise, and
no good fit using this more complex model cound be found.

It can be seen from Table~\ref{tab_params} that the radius of the shell in
models which include two point-sources is 5--10\% larger than that obtained
in the shell-only model used by Gaensler et al. (1997). It is also clear
that the point sources are moving outward with the expansion of the shell,
the eastern one apparently at a larger relative rate than the shell. As for
the shell model, we can fit these larger radii by a linear increase, to
obtain an expansion speed of $2300\pm300$~\kms, about 35\% slower than in
the case for the shell alone. However, we note that if multi-parameter fits
like those in the third row of Figure~\ref{fig_models} are carried out for
other epochs, the resulting radii are $\sim$5\% {\em smaller}\ than for the
fits to a shell alone, and the resulting inferred expansion speed is
$\sim3700$~\kms. While we have not carried out the corresponding analysis
here, we note that Staveley-Smith et al. (1993a) fitted the radio emission
from SNR 1987A with a thick spherical shell of outer-to-inner radius ratio
1.25. This results in a shell diameter about 10\% greater than for the
thin-shell fit, consistent with the range of possible radii considered here.

We therefore conclude that the radius of the remnant as determined
from fitting to the $u-v$ data is uncertain by $\sim$10\%, and that the
resulting expansion velocity is uncertain by $\sim$30\%. The main
conclusions from earlier results --- that the radio emission is
originating from a region within the equatorial ring, and that the
material producing this emission was initially moving rapidly but now
has a very low rate of expansion --- are unchanged, even when the
assumption of spherical symmetry is relaxed.

\subsection{Comparison With Other Wavelengths and Discussion}


\begin{figure}[htb]
\begin{center}
\psfig{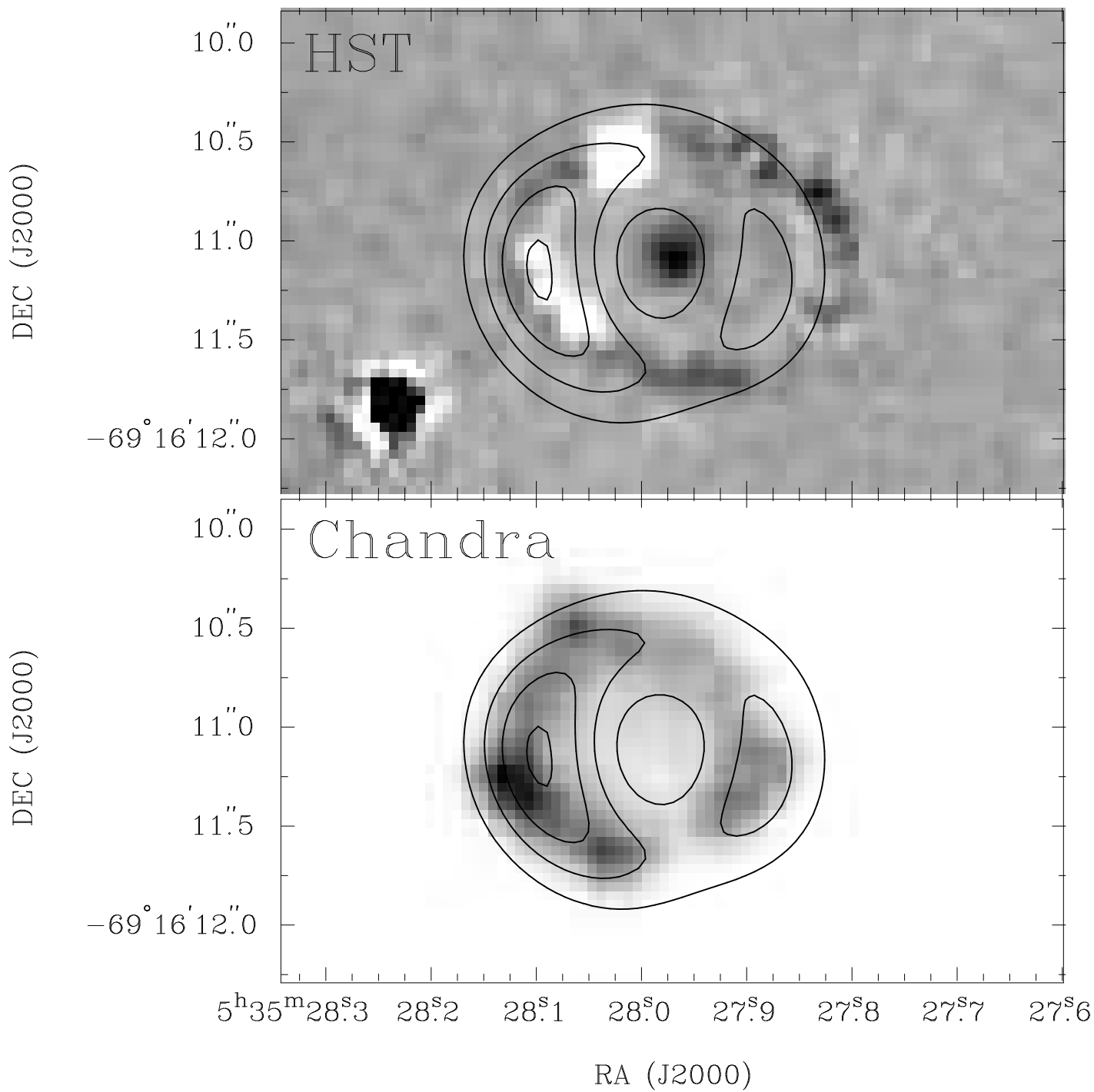}
\caption{Multi-wavelength comparisons of SNR~1987A.  The upper panel
overlays {\em HST}\ and ATCA observations of SNR~1987A: the false-colour image
represents the difference of {\em HST}\ WFPC2 images in the F6565N filter,
obtained by subtracting an image taken in 1998 February from a similar image
taken in 1999 April (Lawrence et al. 2000), while the contours correspond to
the super-resolved 2000.8-epoch ATCA image, using contour levels of 1.1,
2.2, 3.3 and 4.4~mJy~beam$^{-1}$.  The lower panel overlays {\em Chandra}\
and ATCA observations --- the false-colour represents a super-resolved {\em
Chandra}\ ACIS-S image covering the energy range 0.3--8~keV at epoch 1999.8
(Burrows et al.  2000), while the contours indicate
radio data as in the upper panel.}
\label{fig_overlay}
\end{center}
\end{figure}

In Figure~\ref{fig_overlay} we compare our 9~GHz ATCA data to recent
observations with {\em HST}\ and with {\em Chandra}.  In the upper panel,
the super-resolved ATCA image from epoch 2000.8 is compared to the
difference-image of optical emission around the supernova produced by
Lawrence et al. (2000). The HST image has been registered on the radio frame
with an accuracy of better than $0\farcs1$ using the position of the central
star from Reynolds et al. (1995). It shows both the fading equatorial ring,
and and at least seven hotspots on this ring, all brightening as the
supernova shock begins to interact with dense circumstellar material. It can
be seen from this optical/radio comparison that the conclusions made by
Gaensler et al. (1997) are maintained: the bright radio lobes align with the
major axis of the optical ring, with the brighter eastern lobe clearly more
distant from the central star than the fainter western lobe.  Within the
uncertainties, the eastern radio lobe is coincident with some of the optical
hotspots (but not the brighest one), just inside the optical ring.  In the
west, the optical emission lies outside the radio lobe, close to the lowest
radio contour in Figure~\ref{fig_overlay}. In fact, there is a good
correspondence of the radio emission with optical emission {\em within}\ the
optical ring (not visible in Figure~\ref{fig_overlay}) corresponding to the
reverse-shock, as reported by Garnavich, Kirshner \& Challis (1999).  Just
as in the radio data, two optical lobes are seen, one to the east and one to
the west of the supernova site. Also similar to the case in the radio, the
eastern optical lobe is brighter than and further from the supernova than
the western lobe.


The lower panel of Figure~\ref{fig_overlay} compares radio emission with a
super-resolved image obtained by {\em Chandra}\ at epoch 1999.8 (Burrows et
al.  2000). The X-ray and radio emission from the supernova remnant both
take the form of limb-brightened shells; the images were aligned by placing
the estimated centre of the X-ray remnant on the Reynolds et al. (1995)
position (D. Burrows, private communication) and hence is less accurate than
the optical--radio alignment.  While there is a good correspondence between
the brightest regions in each waveband, the X-ray maxima appear to lie
outside the radio maxima. This is perhaps surprising since theoretical
models (e.g. Borkowski, Blondin \& McCray 1997) suggest that the radio
emission is generated just inside the outer shock whereas the X-ray emission is
generated by a reverse shock compressing and heating the denser ejecta as it
propagates inwards. It is worth noting, however, that X-ray emission lying
outside the radio emission and interpreted as coming from the outer shock
has been detected in the SNR 1E 0102.2$-$7219 by Gaetz et al. (2000).

The two-lobed radio morphology seen for SNR~1987A was apparent at least as
early as 1992 (Figure~\ref{fig_movie_super}; Staveley-Smith et al 1993a),
while optical hotspots did not begin to appear on the eastern side of the
optical ring until 1995 and on the western side until 1998 (Lawrence et
al. 2000). Furthermore, the earliest-appearing and brightest optical hotspot
does not coincide in position angle with either of the two main radio lobes,
nor with the brightest X-ray emission seen by {\em Chandra}\
(Figure~\ref{fig_overlay}; Burrows et al. 2000).  We therefore argue that
the more rapid rate in the increase of radio emission beginning around day
3000, as reported by Ball et al. (2001) and confirmed in Section~2 above, is
not related to the appearance of optical hotspots seen at around the same
time.  While Ball et al. (2001) have pointed out that the various optical
hotspots turned on at about the right time for them to have been produced by
the arrival of the radio-producing shock(s) in these regions, it seems clear
that this had little effect on the radio morphology of the remnant. The
best explanation for deviations from spherical symmetry in the radio
morphology of SNR~1987A still seems to be that they result simply from
regions of enhanced emission in the equatorial plane of the progenitor
system, where circumstellar gas is expected to be both densest and closest
to the progenitor star (cf. Gaensler et al. 1997).  The fact that at both
radio and optical wavelengths the eastern lobe is both brighter and more
distant from the explosion site than the western lobe may represent an
asymmetry in the distribution of ejecta.

It is notable that the emission underlying the radio hotspots is well
described by a spherical shell and that, in all of the models, the
flux density of the shell dominates the total flux density. This is
surprising given the strong equatorial enhancement evident in the
optical data and implied for the circumstellar gas. One might expect
eventually to see a faster expansion of the radio remnant in the polar
directions (Blondin, Lundqvist \& Chevalier 1996), but there is
currently no evidence for this. Part of the radio emission from the
lobes may be attributed to an equatorial enhancement but, since such
an enhancement would be symmetric, it cannot account for all of the
emission seen from the brighter eastern lobe. It is worth noting that
the shell of SN 1993J also appears to be quite spherical (Marcaide et
al. 1997, Bartel et al. 2000).

\section{Future Prospects}
Results at all wavelengths suggest that there is an increasing interaction
between the expanding ejecta and the circumstellar material. Hotspots around
or just inside the emission-line ring are certainly becoming more numerous
and prominent in the optical band. Neither the radio imaging nor the X-ray
imaging has sufficient resolution to separately identify hot spots. However,
the modelling of the radio $u-v$ data and the possible short-term
fluctuations in the rise of the high-frequency radio flux densities suggest
that compact radio hotspots do exist.  The overall morphology and the
time-evolution of the radio emission suggest that there is no detailed
correspondence of the radio hotspots with the hotspots seen in the optical
data and, in fact, that the radio hotspots are much more numerous and
widespread. 

We have shown that the rate of increase of the radio emission from SNR 1987A
has slowly increased over the past few years.  It is possible that the rate
of increase of the radio (and other) emission will dramatically increase when
significant amounts of ejecta begin to interact with the dense circumstellar
gas of the inner ring. Extrapolation of the radii and expansion speeds
resulting model fits to the radio data suggest that this will happen in
$2004\pm2$. The ATCA is presently being upgraded for observations in the
12~mm and 3~mm bands, with an expected commissioning date of mid-2002. While
the radio remnant is unlikely to be detectable at 3~mm, we expect 12~mm
observations to provide increased resolution. Extrapolating the present flux
density increase and the currently observed power-law spectrum, we expect a
flux density at 20 GHz in mid-2002 of about 17 mJy. The diffraction limited
half-power beamwidth at 20 GHz will be about 0\farcs4, somewhat less than
the present super-resolved beamwidth at 9~GHz. Even in 2002, it is likely
that super-resolution will be able to be applied to the 20 GHz data, giving
a resolution of about 0\farcs2, and this will certainly be true at later
epochs if the flux density continues to rise. We hope and expect that this
will reveal further detail in the radio images, allowing interesting
comparisons with images obtained at other wavelengths and giving further
insight into the physics of the radio emission process.



\section*{Acknowledgements}


We thank Lewis Ball for making available a pre-publication file of the MOST
flux densities for SNR 1987A and for comments on an earlier version of the
paper, Dave Burrows for providing the {\em Chandra}\ data, and Ben Sugerman
and Peter Garnavich for providing {\em HST}\ data.  We also thank the
referees for helpful suggestions. The Australia Telescope is funded by the
Commonwealth of Australia for operation as a National Facility managed by
CSIRO.  BMG acknowledges the support of NASA through Hubble Fellowship grant
HF-01107.01-98A awarded by the Space Telescope Science Institute, which is
operated by the Association of Universities for Research in Astronomy, Inc.,
for NASA under contract NAS 5--26555.

\section*{References}


\reference
Ball, L., Campbell-Wilson, D., Crawford, D.~F., \& Turtle, A.~J. 1995,
  ApJ, 453, 864

\reference
Ball, L., Crawford, D.~F., Hunstead, R.~W., Klamer, I., \& McIntyre, V.~J.
  2001, ApJ, 549, 599

\reference
Ball, L. \& Kirk, J.~G. 1992, ApJ Lett., 396, L39

\reference
{Baring}, M.~G., {Ellison}, D.~C., {Reynolds}, S.~P., {Grenier}, I.~A., \&
  {Goret}, P. 1999, ApJ, 513, 311

\reference
Bartel, N. et al. 2000, Science, 287, 112 

\reference
Blondin, J.~M., Lundqvist, P. \& Chevalier, R.~A., 1996, ApJ, 472, 257

\reference
Borkowski, K.~J., Blondin, J.~M. \& McCray, R. 1997, ApJ, 476, L31

\reference
Briggs, D.~S. 1994, in { The Restoration of HST Images and Spectra II}, ed.\
  R.~J. Hanisch \& R.~L. White, (Baltimore: Space Telescope Science Institute),
  250

\reference
Burrows, C.~J. {et al.}  1995, ApJ, 452, 680

\reference
{Burrows}, D.~N. {et al.}  2000, ApJ, 543, L149

\reference
Chevalier, R.~A. 1992, Nature, 355, 617

\reference
Chevalier, R.~A. \& Dwarkadas, V.~V. 1995, ApJ, 452, L45

\reference
Chevalier, R.~A. \& Fransson, C. 1987, Nature, 328, 44

\reference
Crotts, A. P.~S., Kunkel, W.~E., \& Heathcote, S.~R. 1995, ApJ, 438,
  724

\reference
Duffy, P., Ball, L., \& Kirk, J.~G. 1995, ApJ, 447, 364

\reference
Gaensler, B.~M., Manchester, R.~N., Staveley-Smith, L., Tzioumis, A.~K.,
  Reynolds, J.~E., \& Kesteven, M.~J. 1997, ApJ, 479, 845

\reference
{Gaetz}, T.~J., {Butt}, Y.~M., {Edgar}, R.~J., {Eriksen}, K.~A., 
{Plucinsky}, P.~P., {Schlegel}, E.~M. \& {Smith}, R.~K. 2000, ApJ, 534, L47

\reference
{Garnavich}, P., {Kirshner}, R., \& {Challis}, P. 1999.
\newblock IAU Circ. 7102

\reference
Gorenstein, P., Hughes, J.~P., \& Tucker, W.~H. 1994, ApJ, 420,
  L25

\reference
Green, D.~A. 2000, { A Catalogue of Galactic Supernova Remnants}, (Cambridge:
  Mullard Radio Astronomy Observatory, Cavendish Laboratory).
\newblock http://www.mrao.cam.ac.uk/surveys/snrs/.

\reference
Hasinger, G., Aschenbach, B., \& Tr\"{u}mper, J. 1996, A\&A, 312,
  L9

\reference
Jauncey, D.~L., Kemball, A., Bartel, N., Shapiro, I.~I., Whitney, A.~R.,
 Rogers, A.~E.~E., Preston, R.~A., \& Clark, T. 1988,
 Nature, 334, 412

\reference
Jones, F.~C. \& Ellison, D.~C. 1991, Space Sci. Rev., 58, 259

\reference
{Lawrence}, S.~S., {Sugerman}, B.~E., {Bouchet}, P., {Crotts}, A. P.~S.,
  {Uglesich}, R., \& {Heathcote}, S. 2000, ApJ, 537, L123

\reference
Marcaide, J.~E. et al. 1997, ApJ, 486, L31

\reference
Montes, M.~J., van Dyk, S.~D., Weiler, K.~W., Sramek, R.~A., \& Panagia, N.
1998, ApJ, 506, 874

\reference
Montes, M.~J., Weiler, K.~W., Van Dyk,
S.~D., Panagia, N., Lacey, C.~K., Sramek, R.~A., \& Park, R. 2000, ApJ, 
532, 1124

\reference
Plait, P.~C., Lundqvist, P., Chevalier, R.~A., \& Kirshner, R.~P. 1995,
  ApJ, 439, 730

\reference
Pun, C. S.~J., Sonneborn, G., Bowers, C., Gull, T., Heap, S., Kimble, R.,
  Maran, S., \& Woodgate, B. 1997.
\newblock IAU Circ. 6665

\reference
Reynolds, J.~E. {et al.}  1995, A\&A, 304, 116

\reference
{Reynolds}, S.~P. \& {Ellison}, D.~C. 1992, ApJ, 399, L75

\reference
Spyromilio, J., Stathakis, R.~A. \& Meurer, G.~R. 1993, MNRAS, 263, 530

\reference
Spyromilio, J. 1994, MNRAS, 266, 61

\reference Staveley-Smith, L., Manchester, R.~N., Kesteven, M.~J.,
Campbell-Wilson, D., Crawford, D.~F., Turtle, A.~J., Reynolds, J.~E.,
Tzioumis, A.~K., Killeen, N.~E.~B. \& Jauncey, D.~L.  1992, Nature, 355,
147

\reference
Staveley-Smith, L., Briggs, D.~S., Rowe, A. C.~H., Manchester, R.~N., Reynolds,
  J.~E., Tzioumis, A.~K., \& Kesteven, M.~J. 1993a, Nature, 366, 136

\reference
Staveley-Smith, L., Manchester, R.~N., Kesteven, M.~J., Tzioumis, A.~K., \&
  Reynolds, J.~E. 1993b, PASA, 10, 331.

\reference
Storey, M.~C. \& Manchester, R.~N. 1987, Nature, 329, 421

\reference
Turtle, A.~J. {et al.}  1987, Nature, 327, 38

\reference
van Dyk, S.~D.,
Weiler, K.~W., Sramek, R.~A., Rupen, M.~P., \& Panagia, N. 1994, ApJ, 
432, L115

\reference
{Weiler}, K.~W., {van Dyk}, S.~D., {Montes}, M.~J., {Panagia}, N., \& {Sramek},
  R.~A. 1998, ApJ, 500, 51






\clearpage
\section*{Appendix}

Table~\ref{tb:fluxes} gives flux densities measured at the four ATCA
frequencies over the day range 2000 - 5100 (MJD 48449 to 51949). These are
calibrated relative to the primary flux calibrator, PKS B1934-638, using
standard techniques. This table, also including flux density
measurements for the nearby unresolved source J0536-6918 and more recent
measurements, is available at
http://www.atnf.csiro.au/research/pulsar/snr/sn1987a/.
 
\begin{table}[htb]
\begin{center}
\caption{ATCA flux density measurements for SNR 1987A}
\label{tb:fluxes}
{\small
\begin{tabular}{cc|cc|cc|cc} \hline
\multicolumn{2}{c}{1.4 GHz} & \multicolumn{2}{c}{2.4 GHz} & 
\multicolumn{2}{c}{4.8 GHz} & \multicolumn{2}{c}{8.6 GHz} \\ \hline
 Day & Flux Density & Day & Flux Density & Day & Flux Density & Day & 
Flux Density \\
Number & (mJy) & Number  & (mJy) & Number  & (mJy) & Number  & (mJy) \\ \hline
 1243.9 & $  0.00\pm 1.20$ & 1198.9 & $  0.00\pm 1.00$ &  1243.6  & $ 0.00\pm 0.25$  & 1388.1 &$  1.33\pm 0.26$ \\
 1270.7 & $  2.92\pm 1.20$ & 1314.5 & $  3.27\pm 1.00$ &  1269.7  & $ 0.66\pm 0.25$  & 1432.2 &$  1.12\pm 0.26$ \\
 1386.4 & $  4.87\pm 1.21$ & 1386.0 & $  3.45\pm 1.01$ &  1305.5  & $ 1.15\pm 0.25$  & 1500.5 &$  1.74\pm 0.26$ \\
 1490.0 & $  7.61\pm 1.22$ & 1517.9 & $  6.13\pm 1.02$ &  1333.7  & $ 1.49\pm 0.26$  & 1515.0 &$  2.34\pm 0.28$ \\
 1517.9 & $  8.29\pm 1.23$ & 1517.7 & $  6.20\pm 1.02$ &  1366.3  & $ 1.53\pm 0.26$  & 1583.8 &$  2.62\pm 0.28$ \\
 1517.7 & $  9.55\pm 1.23$ & 1595.8 & $  5.55\pm 1.01$ &  1385.2  & $ 1.96\pm 0.26$  & 1593.8 &$  2.21\pm 0.27$ \\
 1595.8 & $ 10.15\pm 1.24$ & 1636.7 & $  8.71\pm 1.03$ &  1401.4  & $ 2.17\pm 0.26$  & 1662.5 &$  2.92\pm 0.29$ \\
 1636.7 & $ 15.52\pm 1.29$ & 1660.5 & $  9.15\pm 1.04$ &  1402.4  & $ 2.15\pm 0.26$  & 1661.5 &$  2.50\pm 0.28$ \\
 1660.5 & $ 15.52\pm 1.29$ & 1788.8 & $ 10.67\pm 1.05$ &  1403.4  & $ 2.25\pm 0.27$  & 1786.6 &$  3.92\pm 0.32$ \\
 1788.8 & $ 20.59\pm 1.35$ & 1850.1 & $ 13.07\pm 1.07$ &  1404.3  & $ 2.12\pm 0.26$  & 1852.1 &$  4.37\pm 0.33$ \\
 1850.1 & $ 24.80\pm 1.41$ & 1879.2 & $ 15.03\pm 1.10$ &  1405.3  & $ 2.13\pm 0.26$  & 1878.2 &$  4.14\pm 0.32$ \\
 1879.2 & $ 24.04\pm 1.40$ & 1969.7 & $ 17.95\pm 1.14$ &  1407.1  & $ 2.06\pm 0.26$  & 1947.8 &$  4.70\pm 0.34$ \\
 1969.7 & $ 27.51\pm 1.46$ & 2067.7 & $ 19.00\pm 1.15$ &  1408.3  & $ 2.57\pm 0.27$  & 1969.5 &$  4.93\pm 0.35$ \\
 2068.4 & $ 33.15\pm 1.56$ & 2313.7 & $ 23.25\pm 1.22$ &  1409.3  & $ 2.53\pm 0.27$  & 1986.4 &$  4.64\pm 0.34$ \\
 2313.7 & $ 38.87\pm 1.67$ & 2427.3 & $ 23.37\pm 1.22$ &  1410.3  & $ 2.66\pm 0.27$  & 2003.4 &$  5.09\pm 0.36$ \\
 2427.3 & $ 39.96\pm 1.70$ & 2505.2 & $ 25.91\pm 1.27$ &  1431.2  & $ 2.29\pm 0.27$  & 2067.5 &$  5.48\pm 0.37$ \\
 2505.2 & $ 43.23\pm 1.77$ & 2549.2 & $ 26.56\pm 1.28$ &  1446.2  & $ 2.63\pm 0.27$  & 2092.4 &$  5.92\pm 0.39$ \\
 2549.2 & $ 43.83\pm 1.78$ & 2680.7 & $ 28.40\pm 1.31$ &  1460.1  & $ 3.03\pm 0.28$  & 2142.2 &$  5.43\pm 0.37$ \\
 2680.7 & $ 46.20\pm 1.83$ & 2774.5 & $ 29.92\pm 1.34$ &  1516.9  & $ 2.83\pm 0.27$  & 2195.7 &$  4.20\pm 0.33$ \\
 2774.5 & $ 49.26\pm 1.90$ & 2773.7 & $ 30.40\pm 1.35$ &  1524.9  & $ 2.74\pm 0.27$  & 2261.8 &$  6.04\pm 0.39$ \\
 2773.7 & $ 48.60\pm 1.89$ & 2826.7 & $ 28.40\pm 1.31$ &  1586.8  & $ 2.97\pm 0.28$  & 2299.9 &$  7.14\pm 0.44$ \\
 2857.7 & $ 50.90\pm 1.94$ & 2857.7 & $ 29.30\pm 1.33$ &  1594.8  & $ 4.50\pm 0.31$  & 2299.7 &$  5.80\pm 0.38$ \\
 2874.3 & $ 48.40\pm 1.88$ & 2874.3 & $ 25.20\pm 1.25$ &  1635.6  & $ 5.31\pm 0.33$  & 2313.7 &$  6.92\pm 0.43$ \\
 2919.2 & $ 53.77\pm 2.01$ & 2919.2 & $ 33.11\pm 1.41$ &  1662.5  & $ 5.02\pm 0.32$  & 2320.7 &$  6.90\pm 0.43$ \\
 2919.3 & $ 51.38\pm 1.95$ & 2919.3 & $ 34.20\pm 1.43$ &  1746.8  & $ 6.04\pm 0.35$  & 2374.7 &$  5.38\pm 0.37$ \\
 2975.8 & $ 54.97\pm 2.04$ & 2975.8 & $ 33.55\pm 1.42$ &  1786.8  & $ 7.44\pm 0.39$  & 2403.5 &$  7.14\pm 0.44$ \\
 2975.7 & $ 48.40\pm 1.88$ & 2975.7 & $ 32.50\pm 1.40$ &  1852.1  & $ 8.16\pm 0.41$  & 2426.4 &$  7.68\pm 0.46$ \\
 3001.7 & $ 48.80\pm 1.89$ & 3001.7 & $ 33.11\pm 1.41$ &  1878.2  & $ 7.06\pm 0.38$  & 2462.7 &$  6.50\pm 0.41$ \\
 3000.7 & $ 56.20\pm 2.07$ & 3072.5 & $ 34.60\pm 1.44$ &  1947.8  & $ 8.27\pm 0.41$  & 2550.2 &$  7.47\pm 0.45$ \\
 3072.5 & $ 61.70\pm 2.21$ & 3139.7 & $ 36.60\pm 1.49$ &  1969.5  & $ 9.11\pm 0.44$  & 2572.0 &$  5.93\pm 0.39$ \\
 3072.5 & $ 57.20\pm 2.09$ & 3176.7 & $ 35.90\pm 1.47$ &  1986.4  & $ 8.44\pm 0.42$  & 2579.8 &$  6.92\pm 0.43$ \\
 3139.7 & $ 61.00\pm 2.19$ & 3202.7 & $ 39.20\pm 1.54$ &  2003.4  & $ 8.72\pm 0.43$  & 2579.7 &$  6.50\pm 0.41$ \\
 3176.7 & $ 59.10\pm 2.14$ & 3277.7 & $ 41.60\pm 1.60$ &  2067.2  & $ 9.42\pm 0.45$  & 2628.0 &$  9.00\pm 0.51$ \\
\hline 
\end{tabular}}
\end{center} 	
\end{table}    

\addtocounter{table}{-1}

\begin{table}[htb]
\begin{center}
\caption{ -- {\it continued}}
{\small
\begin{tabular}{cc|cc|cc|cc} \hline
\multicolumn{2}{c}{1.4 GHz} & \multicolumn{2}{c}{2.4 GHz} & 
\multicolumn{2}{c}{4.8 GHz} & \multicolumn{2}{c}{8.6 GHz} \\ \hline
 Day & Flux Density & Day & Flux Density & Day & Flux Density & Day & 
Flux Density \\
Number & (mJy) & Number  & (mJy) & Number  & (mJy) & Number  & (mJy) \\ \hline
 3202.7 & $ 64.20\pm 2.27$ & 3325.7 & $ 41.40\pm 1.59$ &  2092.4  & $ 9.94\pm 0.47$  & 2627.7 &$  7.80\pm 0.46$ \\
 3277.7 & $ 66.20\pm 2.32$ & 3414.8 & $ 44.10\pm 1.66$ &  2142.5  & $ 8.56\pm 0.42$  & 2647.7 &$  6.10\pm 0.39$ \\
 3325.7 & $ 67.10\pm 2.34$ & 3455.7 & $ 44.80\pm 1.68$ &  2195.7  & $ 9.10\pm 0.44$  & 2754.8 &$  6.26\pm 0.40$ \\
 3414.8 & $ 71.80\pm 2.47$ & 3515.4 & $ 45.80\pm 1.70$ &  2261.8  & $10.86\pm 0.50$  & 2753.7 &$  5.40\pm 0.37$ \\
 3455.7 & $ 71.80\pm 2.47$ & 3579.2 & $ 48.50\pm 1.77$ &  2299.9  & $11.50\pm 0.52$  & 2774.5 &$  9.22\pm 0.52$ \\
 3515.4 & $ 73.30\pm 2.51$ & 3633.1 & $ 48.00\pm 1.75$ &  2299.7  & $11.60\pm 0.53$  & 2826.7 &$  9.80\pm 0.55$ \\
 3579.2 & $ 77.50\pm 2.62$ & 3679.0 & $ 49.40\pm 1.79$ &  2374.7  & $10.82\pm 0.50$  & 2919.2 &$  9.44\pm 0.53$ \\
 3633.1 & $ 79.10\pm 2.66$ & 3714.0 & $ 50.10\pm 1.81$ &  2403.5  & $12.42\pm 0.56$  & 2919.3 &$ 11.30\pm 0.62$ \\
 3679.0 & $ 79.60\pm 2.67$ & 3744.7 & $ 48.50\pm 1.77$ &  2462.7  & $12.80\pm 0.57$  & 2975.8 &$  9.11\pm 0.52$ \\
 3714.0 & $ 79.30\pm 2.66$ & 3771.6 & $ 49.90\pm 1.80$ &  2511.0  & $11.98\pm 0.54$  & 2975.7 &$  8.30\pm 0.48$ \\
 3744.7 & $ 84.40\pm 2.80$ & 3833.6 & $ 50.10\pm 1.81$ &  2572.0  & $13.99\pm 0.61$  & 3000.7 &$  6.60\pm 0.41$ \\
 3771.6 & $ 81.90\pm 2.73$ & 3900.3 & $ 52.60\pm 1.87$ &  2579.8  & $13.62\pm 0.60$  & 3001.7 &$  9.23\pm 0.52$ \\
 3833.6 & $ 84.20\pm 2.80$ & 3945.4 & $ 52.20\pm 1.86$ &  2579.7  & $13.10\pm 0.58$  & 3072.5 &$ 10.40\pm 0.58$ \\
 3900.3 & $ 86.00\pm 2.85$ & 3987.1 & $ 52.40\pm 1.86$ &  2628.0  & $13.53\pm 0.60$  & 3073.6 &$ 11.40\pm 0.62$ \\
 3945.4 & $ 88.10\pm 2.90$ & 4015.1 & $ 55.30\pm 1.94$ &  2627.7  & $14.10\pm 0.62$  & 3110.6 &$ 10.20\pm 0.57$ \\
 3987.1 & $ 93.80\pm 3.06$ & 4058.5 & $ 58.30\pm 2.01$ &  2647.7  & $12.90\pm 0.57$  & 3176.7 &$  9.80\pm 0.55$ \\
 4015.1 & $ 89.30\pm 2.94$ & 4100.8 & $ 58.30\pm 2.01$ &  2754.8  & $14.17\pm 0.62$  & 3202.7 &$ 12.60\pm 0.68$ \\
 4058.5 & $ 87.70\pm 2.89$ & 4169.5 & $ 61.40\pm 2.10$ &  2753.7  & $13.61\pm 0.60$  & 3277.7 &$ 12.10\pm 0.65$ \\
 4100.8 & $ 96.70\pm 3.14$ & 4222.5 & $ 59.20\pm 2.04$ &  2774.5  & $14.63\pm 0.64$  & 3325.7 &$ 11.70\pm 0.64$ \\
 4169.5 & $ 99.60\pm 3.22$ & 4291.4 & $ 64.70\pm 2.18$ &  2773.7  & $14.10\pm 0.62$  & 3414.8 &$ 11.50\pm 0.63$ \\
 4222.5 & $ 97.50\pm 3.16$ & 4373.1 & $ 62.30\pm 2.12$ &  2826.7  & $14.50\pm 0.63$  & 3436.7 &$ 15.70\pm 0.82$ \\
 4291.4 & $104.50\pm 3.36$ & 4424.0 & $ 67.10\pm 2.25$ &  2919.2  & $16.47\pm 0.70$  & 3455.7 &$ 14.20\pm 0.75$ \\
 4373.1 & $108.30\pm 3.46$ & 4460.9 & $ 68.90\pm 2.30$ &  2919.3  & $16.10\pm 0.69$  & 3485.5 &$ 16.10\pm 0.84$ \\
 4424.0 & $107.40\pm 3.44$ & 4539.8 & $ 75.50\pm 2.48$ &  2975.8  & $15.28\pm 0.66$  & 3512.4 &$ 16.50\pm 0.86$ \\
 4460.9 & $113.20\pm 3.60$ & 4571.7 & $ 69.50\pm 2.31$ &  2975.7  & $16.50\pm 0.71$  & 3515.4 &$ 14.60\pm 0.77$ \\
 4539.8 & $112.60\pm 3.58$ & 4684.9 & $ 75.20\pm 2.47$ &  3000.7  & $15.10\pm 0.65$  & 3579.2 &$ 15.50\pm 0.81$ \\
 4571.7 & $116.10\pm 3.68$ & 4728.9 & $ 74.20\pm 2.44$ &  3001.7  & $16.84\pm 0.72$  & 3633.1 &$ 14.90\pm 0.79$ \\
 4684.9 & $121.20\pm 3.83$ & 4768.1 & $ 77.10\pm 2.52$ &  3072.5  & $17.80\pm 0.75$  & 3679.0 &$ 16.30\pm 0.85$ \\
 4728.9 & $121.40\pm 3.83$ & 4799.9 & $ 77.40\pm 2.53$ &  3176.7  & $16.90\pm 0.72$  & 3714.0 &$ 15.90\pm 0.83$ \\
 4768.1 & $131.30\pm 4.12$ & 4837.9 & $ 83.80\pm 2.71$ &  3202.7  & $20.20\pm 0.85$  & 3744.7 &$ 16.30\pm 0.85$ \\
 4799.9 & $124.90\pm 3.93$ & 4850.8 & $ 77.80\pm 2.54$ &  3277.7  & $21.60\pm 0.90$  & 3771.6 &$ 17.90\pm 0.93$ \\
 4837.9 & $126.40\pm 3.98$ & 4870.7 & $ 77.90\pm 2.54$ &  3325.7  & $20.70\pm 0.86$  & 3833.6 &$ 17.70\pm 0.92$ \\
 4850.8 & $128.70\pm 4.04$ & 4936.5 & $ 81.00\pm 2.63$ &  3414.8  & $21.30\pm 0.89$  & 3900.3 &$ 17.10\pm 0.89$ \\
 4870.7 & $127.00\pm 3.99$ & 4997.3 & $ 83.60\pm 2.70$ &  3455.7  & $23.10\pm 0.96$  & 3945.4 &$ 17.90\pm 0.93$ \\
 4936.5 & $133.40\pm 4.18$ & 5024.8 & $ 83.00\pm 2.68$ &  3515.4  & $24.80\pm 1.02$  & 3987.1 &$ 18.90\pm 0.98$ \\
 4997.3 & $136.90\pm 4.28$ & 5050.1 & $ 86.20\pm 2.77$ &  3579.2  & $25.40\pm 1.05$  & 4015.1 &$ 19.40\pm 1.00$ \\
 5024.8 & $135.00\pm 4.22$ & 5093.0 & $ 86.60\pm 2.78$ &  3633.1  & $25.10\pm 1.03$  & 4100.8 &$ 19.60\pm 1.01$ \\
 5050.1 & $135.90\pm 4.25$ &        &                  &  3679.0  & $24.50\pm 1.01$  & 4222.5 &$ 19.10\pm 0.99$ \\
 5093.0 & $143.40\pm 4.47$ &        &                  &  3714.0  & $24.50\pm 1.01$  & 4291.4 &$ 21.00\pm 1.08$ \\
        &                  &        &                  &  3744.7  & $27.00\pm 1.11$  & 4373.1 &$ 21.20\pm 1.09$ \\
        &                  &        &                  &  3771.6  & $27.10\pm 1.11$  & 4424.0 &$ 24.90\pm 1.27$ \\
        &                  &        &                  &  3833.6  & $26.50\pm 1.09$  & 4460.9 &$ 25.50\pm 1.30$ \\
\hline 
\end{tabular}}
\end{center} 	
\end{table}    

\addtocounter{table}{-1}

\begin{table}[htb]
\begin{center}
\caption{ -- {\it continued}}
{\small
\begin{tabular}{cc|cc|cc|cc} \hline
\multicolumn{2}{c}{1.4 GHz} & \multicolumn{2}{c}{2.4 GHz} & 
\multicolumn{2}{c}{4.8 GHz} & \multicolumn{2}{c}{8.6 GHz} \\ \hline
 Day & Flux Density & Day & Flux Density & Day & Flux Density & Day & 
Flux Density \\
Number & (mJy) & Number  & (mJy) & Number  & (mJy) & Number  & (mJy) \\ \hline
	& 	           &	    &                  &  3900.3  & $29.20\pm 1.19$  & 4539.8 &$ 25.00\pm 1.27$ \\
	& 		   &	    &                  &  3945.4  & $29.80\pm 1.22$  & 4571.7 &$ 24.40\pm 1.25$ \\
        &                  &        &                  &  3987.1  & $31.90\pm 1.30$  & 4684.9 &$ 26.70\pm 1.36$ \\
        &                  &        &                  &  4015.1  & $30.20\pm 1.23$  & 4728.9 &$ 28.10\pm 1.43$ \\
        &                  &        &                  &  4100.8  & $32.00\pm 1.30$  & 4768.1 &$ 25.70\pm 1.31$ \\
        &                  &        &                  &  4222.5  & $32.00\pm 1.30$  & 4799.9 &$ 24.80\pm 1.26$ \\
	& 		   &	    &                  &  4291.4  & $35.30\pm 1.43$  & 4837.9 &$ 31.50\pm 1.59$ \\
	& 		   &	    &                  &  4424.0  & $34.10\pm 1.39$  & 4850.8 &$ 27.90\pm 1.42$ \\
	& 		   &	    &                  &  4460.9  & $38.60\pm 1.56$  & 4870.7 &$ 29.90\pm 1.52$ \\
	&		   &	    &                  &  4539.8  & $37.60\pm 1.52$  & 4936.5 &$ 30.20\pm 1.53$ \\
	&		   &	    &                  &  4571.7  & $39.60\pm 1.60$  & 4997.3 &$ 27.50\pm 1.40$ \\
	&		   &	    &                  &  4684.9  & $40.10\pm 1.62$  & 5024.8 &$ 31.90\pm 1.61$ \\
	&		   &	    &                  &  4728.9  & $43.70\pm 1.77$  & 5050.1 &$ 31.90\pm 1.61$ \\
	&		   &	    &                  &  4768.1  & $42.70\pm 1.73$  & 5093.0 &$ 36.00\pm 1.82$ \\
	&		   &	    &                  &  4799.9  & $39.70\pm 1.61$  &	      &                 \\
	&		   &	    &                  &  4837.9  & $43.60\pm 1.76$  &	      &                 \\
	&		   &	    &                  &  4850.8  & $45.50\pm 1.84$  &	      &                 \\
	&		   &	    &                  &  4870.7  & $47.20\pm 1.90$  &	      &		        \\
	&		   &	    &                  &  4936.5  & $48.10\pm 1.94$  &	      &		        \\
	&		   &	    &                  &  4997.3  & $43.50\pm 1.76$  &	      &		        \\
	&		   &	    &                  &  5024.8  & $47.70\pm 1.92$  &	      &		        \\
	&		   &	    &                  &  5050.1  & $46.40\pm 1.87$  &	      &		        \\
	&		   &	    &                  &  5093.0  & $49.90\pm 2.01$  &        & 	        \\
\hline 
\end{tabular}}
\end{center} 	
\end{table}

\end{document}